\documentclass[aps,prb,twocolumn]{revtex4}

\usepackage{latexsym,amssymb}
\usepackage{graphicx,color,pstricks}
\usepackage{epsfig,epsf,bm}
\begin{document}
%\begin{frontmatter}
	
\title{Magnetic impurities in an altermagnetic metal}

\author{Yu-Li Lee}
\email{yllee@cc.ncue.edu.tw}
%\ead{yllee@cc.ncue.edu.tw}
\affiliation{Department of Physics, National Changhua University of Education, Changhua, Taiwan, R.O.C.}
%\affiliation{organization={Department of Physics, National Changhua University of Education}, city={Changhua}, 
%	state={Taiwan}, country={R.O.C.}}
	
\date{\today}
	
\begin{abstract}
%\abstract{
 We study the physics of dilute magnetic impurities in a two-dimensional altermagnetic metal. For the 
 single impurity case, although the spin degeneracy is broken in an altermagnetic metal, we show that the 
 antiferromagnetic Kondo coupling still flows to the strong coupling regime at low energies in terms of 
 the one-loop renormalization-group equation. Moreover, the Kondo temperature may be enhanced or reduced, 
 depending on the band structure and the electron density. To study the ground-state properties, we employ 
 the variational wavefunction approach. We find that the impurity spin is completely screened at long 
 distances, in contrast to the usual antiferromagnet. The $d$-wave nature of the spin-split Fermi surfaces 
 in an altermagnetic metal is reflected in the correlation between the impurity and conduction electron 
 spins, which exhibits the $C_{4z}$ symmetry of the altermagnet at long distances. The spin correlation 
 decays as $1/r^3$ at long distances, and its amplitude oscillates with four different periods due to the 
 interference between the spin splitting Fermi surfaces. Moreover, the values of these periods depend on 
 the direction of observation. Similar phenomena also occur in the RKKY interaction in an AM metal.
%}
\end{abstract}

%\begin{keyword}
% Altermagnetic metals \sep Kondo problem \sep RKKY interactions
%\end{keyword}
	
%\end{frontmatter}

%\keywords{Altermagnetic metals, Kondo problem, RKKY interactions}

\maketitle
	
%%%%%%%%%%%%%%%%%%%%%%%%%%%%%%%%%%%%%%%%%%%%%%%%%%%%%%%%%%%%%%%%%%%%%%%%%%%%%%%%%%%%%%%%%%%%%%%%%%%%%%%%%
\section{Introduction}

The conventional collinear magnets are divided into two types: the ferromagnets and antiferromagnets, 
depending on whether or not the net magnetization per (magnetic) unit cell vanishes\cite{LLP}. For the 
ferromagnet with one spin sublattice, the net magnetization is nonzero, and thus the time-reversal (T) 
symmetry is broken. It follows that the band structure exhibits momentum-independent spin splitting, 
leading to isotropic $s$-wave spin-split Fermi surfaces. On the other hand, the classical ``N\'eel" 
antiferromagnets with the opposite spin sublattices connected by inversion or translation possess zero 
net magnetization and $T$-invariant spin degenerate bands.

Recently, a new type of collinear magnets, dubbed as the altermagnets, was discovered\cite{HYK1,HYK2,SSJ1}. 
Similar to the classical ``N\'eel" antiferromagnets, the altermagnets have zero net magnetization, and thus 
must contain at least two spin sublattices. However, the opposite spin sublattices in the altermagnet is 
connected by rotation instead of inversion or translation\cite{SSJ2,SSJ3}. Moreover, its band structure 
exhibits $T$-breaking spin splitting, leading to anisotropic $d$ ($g$ or $i$)-wave spin splitting Fermi 
surfaces. The spin splitting in altermagnets is of non-relativistic origin, which is distinct from 
ferromagnetic (FM) and relativistically spin-orbital coupled systems\cite{SSJ2,YLF,LLH,JSZ}. In analogy 
with superconductors, the FM metal is like the $s$-wave superconductor, while the altermagnetic (AM) metal 
is the analog of the $d$-wave superconductor\cite{Mazin}. From {\it ab initio} calculations, several 
materials were predicted to be altermagnets, which may be insulators like MnF$_2$ and MnTe, or metals like 
RuO$_2$\cite{SSJ2,Mazin2,SMA,HYK,FMA}. (However, a recent experiment in terms of spin- and angle-resolved 
photoemission spectroscopy suggested the absence of the AM spin-splitting band structures in RuO$_2$.\cite{exp1})

Recent progress was mainly focused on the aspects of spintronics\cite{DSS,BZZ,SL} and the interplay between 
superconductivity and altermagnetism\cite{SBL,OBL,BV,PAP,ZZW,BBS,ZHN,GHC}. One less studied topic is how
the spin-split and T-breaking bands in an AM metal affect the properties of dilute magnetic impurities at 
low energies. For the Kondo problem in an ordinary Fermi liquid (FL), it is the enhancement of the spin-flip 
scatterings of conduction electrons by magnetic impurities at low energies which results in the complete 
screening of the magnetic impurity at $T=0$. The spin degeneracy (and thus the T symmetry) is important for 
this enhancement, which are manifest in the suppression of the Kondo temperature in the presence of 
ferromagnetism\cite{Martinek1,Martinek2,CSL,Pasupathy}. Especially, in the extreme case of half-metals, i.e., 
the minority-spin electron are completely absent, the screening of the impurity spin is not possible. One 
may wonder how the spin-splitting Fermi surfaces in the AM metal will affect the magnetic impurities at low 
energies. Will the spin-splitting Fermi surfaces in the AM metal suppress the spin-flip scatterings of 
conduction electrons from the magnetic impurities?

In the present work, we try to answer this question in a two-dimensional ($2$D) AM metal in terms of the 
one-loop renormalization group (RG) equation and the variational wavefunction approaches. We find that the 
RG flow of the Kondo coupling is similar to the case in the ordinary FL with isotropic spin-degenerate 
$s$-wave Fermi surfaces. That is, the Kondo coupling flows to the strong coupling regime when it is 
antiferromagnetic (AF), so that the local spin is completely screened by the itinerant electrons. These 
conclusions are justified by a slave-boson mean-field theory. The result we obtained is in contrast to the 
case in the conventional AF metal\cite{AVV}, where the local spin is only partially screened, and is 
consistent with a recent study on the single-impurity Anderson model in terms of the numerical RG\cite{Diniz}. 
We showed that this follows from the $C_{4z}$ symmetry of the AM metal, which leads to the identical 
density of states (DOS) for spin-up and -down bands. In contrast with the claim made in Ref. \onlinecite{Diniz} 
which states that the Kondo temperature is reduced in an AM metal, we indicate that this is an artifact 
resulting from the specific model adopted in it. In general, whether or not the Kondo temperature is 
reduced depends on the band structure as well as the electron density.

The main difference between the Kondo effects in the FL and an AM metal lies at the spin correlation between 
the impurity and conduction electron spins, which measures the spatial extension of the Kondo screening 
cloud. Due to the $d$-wave nature of the spin-split Fermi surfaces in the AM metal, the angular dependence 
of spin correlation has the $C_{4z}$ symmetry. Moreover, the radial dependence of the spin correlation 
exhibits the oscillatory power-law decay, with multi-periods depending on the direction of observation. This 
follows from the interference between the spin splitting Fermi surfaces.

Similar behaviors also appear in the RKKY interaction between two magnetic impurities in a $2$D AM metal. We 
show that the RKKY interaction in the AM metal is of an anisotropic Heisenberg type, in contrast to the 
ordinary FL with an isotropic spin-degenerate $s$-wave Fermi surfaces. Again, this interaction exhibits the 
$C_{4z}$ symmetry following from the $d$-wave spin splitting Fermi surfaces. Moreover, the transverse and 
longitudinal components of the RKKY interaction have distinct magnitudes. At short distance, the transverse 
component is much smaller than the longitudinal one, such that the exchange interaction between the two 
magnetic impurities is of the Ising type. On the other hand, at long distances, the transverse component 
becomes much larger than the longitudinal one, such that the exchange interaction between the two magnetic 
impurities turns into the $XY$ type.

The rest of the work is organized as follows. We first study the Kondo problem in Sec. \ref{kondo} and
present the calculation of the RKKY interaction in Sec. \ref{rkky}. The last section is devoted to a 
conclusive discussion. Three appendices are provided for the details of the calculations in Sec. \ref{kondo}, 
\ref{rkky}, and the slave-boson mean-field theory. 

%%%%%%%%%%%%%%%%%%%%%%%%%%%%%%%%%%%%%%%%%%%%%%%%%%%%%%%%%%%%%%%%%%%%%%%%%%%%%%%%%%%%%%%%%%%%%%%%%%%%%%%%%%%
\section{The Kondo problem}
\label{kondo}
\subsection{The model}

The simplest model describing the $2$D AM metal is given by the following single-particle, two-band Bloch 
Hamiltonian on a square lattice\cite{SSJ1,ZZW}:
\begin{eqnarray}
 \mathcal{H}(\bm{k}) &=& -t_0(\cos{k_x}+\cos{k_y})+t_1\sin{k_x}\sin{k_y}\sigma_3 \nonumber \\
 & & +2t_2(\cos{k_x}-\cos{k_y})\sigma_3 \ , \label{mh1}
\end{eqnarray}
which contains, apart from the common nearest-neighbor hopping term, two spin-dependent hopping terms due 
to the anisotropic exchange interaction in the AM phase. Here, $\sigma_{1,2,3}$ are Pauli matrices in the 
spin space. The resulting electronic band structure is of the form
\begin{eqnarray}
 \epsilon_{\sigma}(\bm{k}) &=& -t_0(\cos{k_x}+\cos{k_y})+\sigma t_1\sin{k_x}\sin{k_y} \nonumber \\
 & & +2\sigma t_2(\cos{k_x}-\cos{k_y}) \ , \label{mh11}
\end{eqnarray}
where $\sigma=\pm$ correspond to spin-up and -down, respectively. Under the $C_{4z}$ rotation, 
$k_x\rightarrow k_y$ and $k_y\rightarrow -k_x$, and we have
\begin{equation}
 \epsilon_{\sigma}(C_4\bm{k})=\epsilon_{-\sigma}(\bm{k}) \ . \label{mh12}
\end{equation}
Since $\sigma_2\mathcal{H}^*(\bm{k})\sigma_2\neq\mathcal{H}(-\bm{k})$ the T symmetry is broken in the AM 
phase.

Both the spin-up and -down bands have a global minimum at the $\Gamma$ point ($\bm{k}=0$), when the parameters 
$t_1$ and $t_2$ satisfy the constraint
\begin{equation}
 \alpha_1^2+4\alpha_2^2<1 \ , \label{amh12}
\end{equation}
where $\alpha_{1/2}=t_{1/2}/t_0$ are dimensionless parameters. Alternatively, one may define 
\begin{equation}
 \alpha_1=\nu\cos{(2\phi)} \ , ~~2\alpha_2=\nu\sin{(2\phi)} \ , \label{amh16}
\end{equation}
with $\nu\geq 0$ and $0\leq\phi\leq\pi$. Then, Eq. (\ref{amh12}) indicates that $\nu<1$. Hereafter, we will
focus on this case.

\begin{figure}
\begin{center}
 \includegraphics[width=0.99\columnwidth]{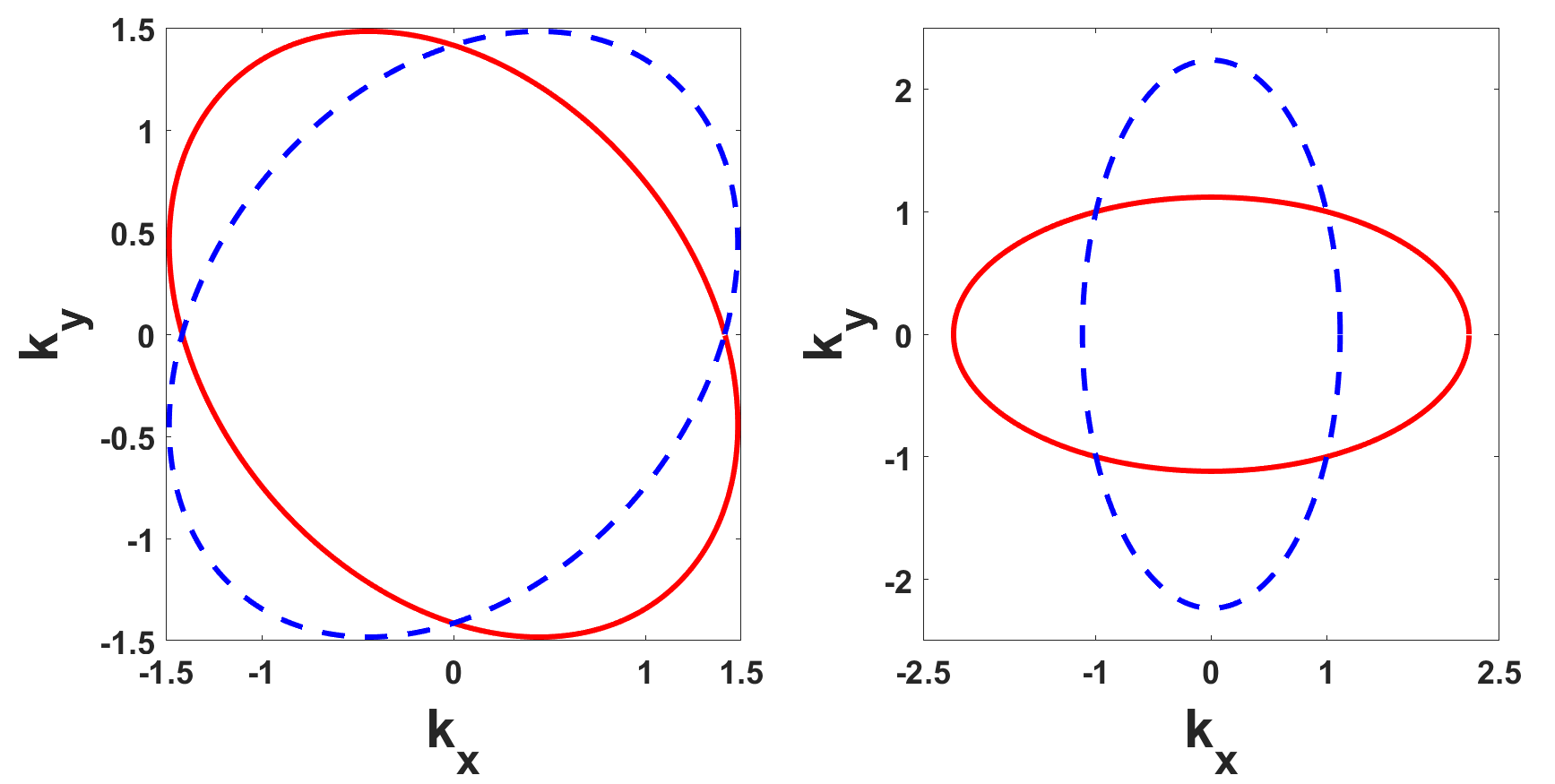}
 \caption{The Fermi surfaces at $\epsilon_F=t_0$ for the $d_{xy}$-wave symmetry (left) with $t_1=0.3t_0$ and 
 $t_2=0$ and $d_{x^2-y^2}$-wave symmetry (right) with $t_2=0.3t_0$ and $t_1=0$. The solid and dashed lines 
 correspond to the spin-up and -down Fermi surfaces, respectively.}
 \label{amfsf1}
\end{center}
\end{figure}

Our working Hamiltonian is $H_0+H_K$ where
\begin{equation}
 H_0=\! \sum_{\bm{k},\sigma=\pm}c^{\dagger}_{\bm{k}\sigma}[\epsilon_{\sigma}(\bm{k})-\mu]c_{\bm{k}\sigma}
 \ , \label{amh1}
\end{equation}
describes the AM metal and
\begin{equation}
 H_K=J\bm{S}\cdot\psi^{\dagger}\bm{\sigma}\psi(0) \ , \label{amh15}
\end{equation}
is the coupling between the conduction electrons and the impurity spin $\bm{S}$. Here $c_{\bm{k}\sigma}$, 
$c^{\dagger}_{\bm{k}\sigma}$ obey the canonical anticommutation relations, $\psi=[\psi_+,\psi_-]^t$, and 
$\psi_{\sigma}(\bm{r})=\frac{1}{\sqrt{V}} \! \sum_{\bm{k}}c_{\bm{k}\sigma}e^{i\bm{k}\cdot\bm{r}}$ with $V$ 
being the volume of the system. Without loss of generality, we take the position of the local spin as the 
origin. 

When $\nu<1$, to perform analytic calculations, we expand $\mathcal{H}(\bm{k})$ around the band minimum 
(the $\Gamma$ point) and employ the following dispersion relations for conduction electrons\cite{SL,BV,PAP}:
\begin{equation}
 \epsilon_{\sigma}(\bm{k})=\frac{\bm{k}^2}{2m}+\sigma[t_1k_xk_y-t_2(k_x^2-k_y^2)]  \ , \label{amh11}
\end{equation}
with $m=1/t_0$. Note that Eq. (\ref{amh11}) still preserves the $C_{4z}$ symmetry. Moreover, the resulting 
Hamiltonian $H_0$ still breaks the T symmetry. Therefore, this approximation keeps the most important 
features of the AM metal. When Eq. (\ref{amh12}) is satisfied, the Fermi surface is closed and consists of 
two branches, which exhibits a $d$-wave nature, as illustrated in Fig. \ref{amfsf1}.

%%%%%%%%%%%%%%%%%%%%%%%%%%%%%%%%%%%%%%%%%%%%%%%%%%%%%%%%%%%%%%%%%%%%%%%%%%%%%%%%%%%%%%%%%%%%%%%%%%%%%%%%%%%
\subsection{The one-loop RG equation}
\label{onerg}

To determine the role of $H_K$ at low energies, we employ the perturbative RG. There are various ways to 
implement the perturbative RG. In all cases, the relevant single-particle Green function of itinerant 
electrons is the local one in the imaginary-time formulation\cite{Hewson}, i.e.,
\begin{eqnarray*}
 \mathcal{G}_{\sigma}^{(0)}(i\omega_n)=\frac{1}{V}\sum_{\bm{k}}\frac{1}{i\omega_n-\epsilon_{\sigma}(\bm{k})+\mu}
 =\! \int^{\Lambda}_{-\Lambda} \! d\epsilon\frac{N_{\sigma}(\epsilon)}{i\omega_n-\epsilon} \ ,
\end{eqnarray*}
where $N_{\sigma}(\epsilon)$ is the DOS for spin-$\sigma$ electrons, with $\epsilon$ being measured 
relative to the Fermi energy $\epsilon_F$, and $\Lambda$ is the UV cutoff in energies. The RG equation for 
$J$ is obtained by integrating out the electronic degrees of freedom within the energy shell 
$[\Lambda/b,\Lambda]$ with $b>1$\cite{Hewson}.

To proceed, we notice that $N_{\sigma}(\epsilon)$ is independent of $\sigma$. This is a generic feature for 
the AM metal, which follows from the $C_{4z}$ symmetry [Eq. (\ref{mh12})], not exclusive for the specific 
model we adopted. It can be shown as follows. From Eq. (\ref{mh12}), $N_{\sigma}(\epsilon)$ 
can be written as
\begin{eqnarray*}
 N_{\sigma}(\epsilon) \! \! &=& \! \! \frac{1}{V}\sum_{\bm{k}\in C}\delta[\epsilon-\epsilon_{\sigma}(\bm{k})]
 =\frac{1}{V}\sum_{\bm{k}\in C}\delta[\epsilon-\epsilon_{\sigma}(C_{4z}\bm{k})] \\
 \! \! &=& \! \! \frac{1}{V}\sum_{\bm{k}\in C}\delta[\epsilon-\epsilon_{-\sigma}(\bm{k})]=N_{-\sigma}(\epsilon)
 \ ,
\end{eqnarray*}
where $C$ denotes the first Brillouin zone which is also invariant under the $C_{4z}$ rotation.

Therefore, the one-loop RG flow for the Kondo coupling is identical to the one in the usual FL. That is\cite{Hewson}, 
\begin{equation}
 \Lambda\frac{\partial g}{\partial\Lambda}=-g^2 \ , \label{amkondo11}
\end{equation}
where $g=2N(0)J$ is the dimensionless coupling constant and $N(0)=N_+(0)=N_-(0)$ is the DOS at the Fermi 
energy. Consequently, $g$ flows to the strong coupling regime at low energies for the AF bare exchange 
interaction ($J_0=J(D)>0$ with $D$ denoting the band width), leading to the screening of the local spin 
in the ground state. In this case, Eq. (\ref{amkondo11}) indicates that $g$ diverges at the energy scale $\Lambda=\Lambda_*=De^{-1/[2N(0)J_0]}$, which may be identified as the Kondo temperature $T_K$. On the 
other hand, $g$ flows to zero at low energies when $J_0<0$, leading to the decoupled local spin and 
itinerant electrons in the ground state.

\begin{figure}
\begin{center}
 \includegraphics[width=0.99\columnwidth]{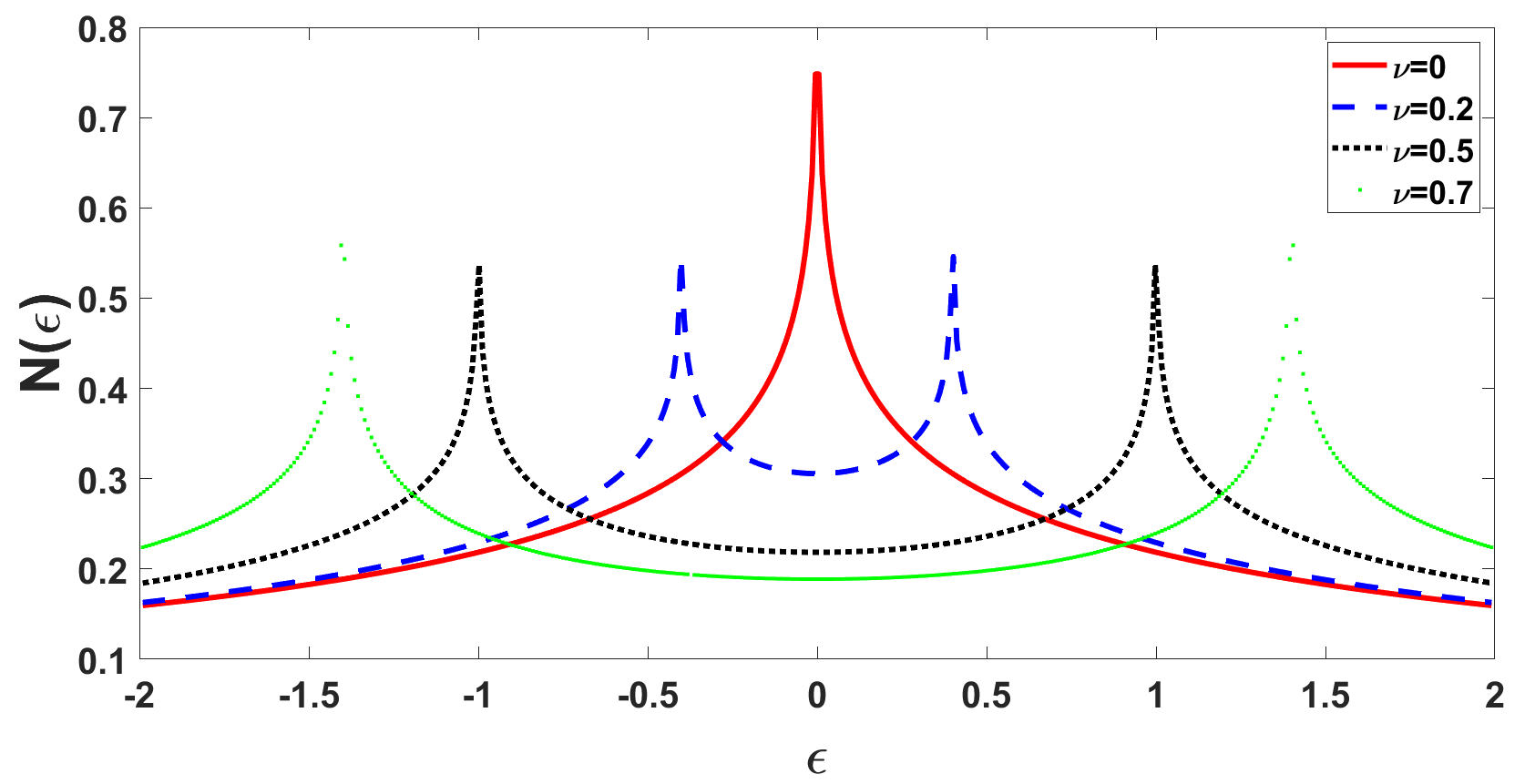}
 \includegraphics[width=0.99\columnwidth]{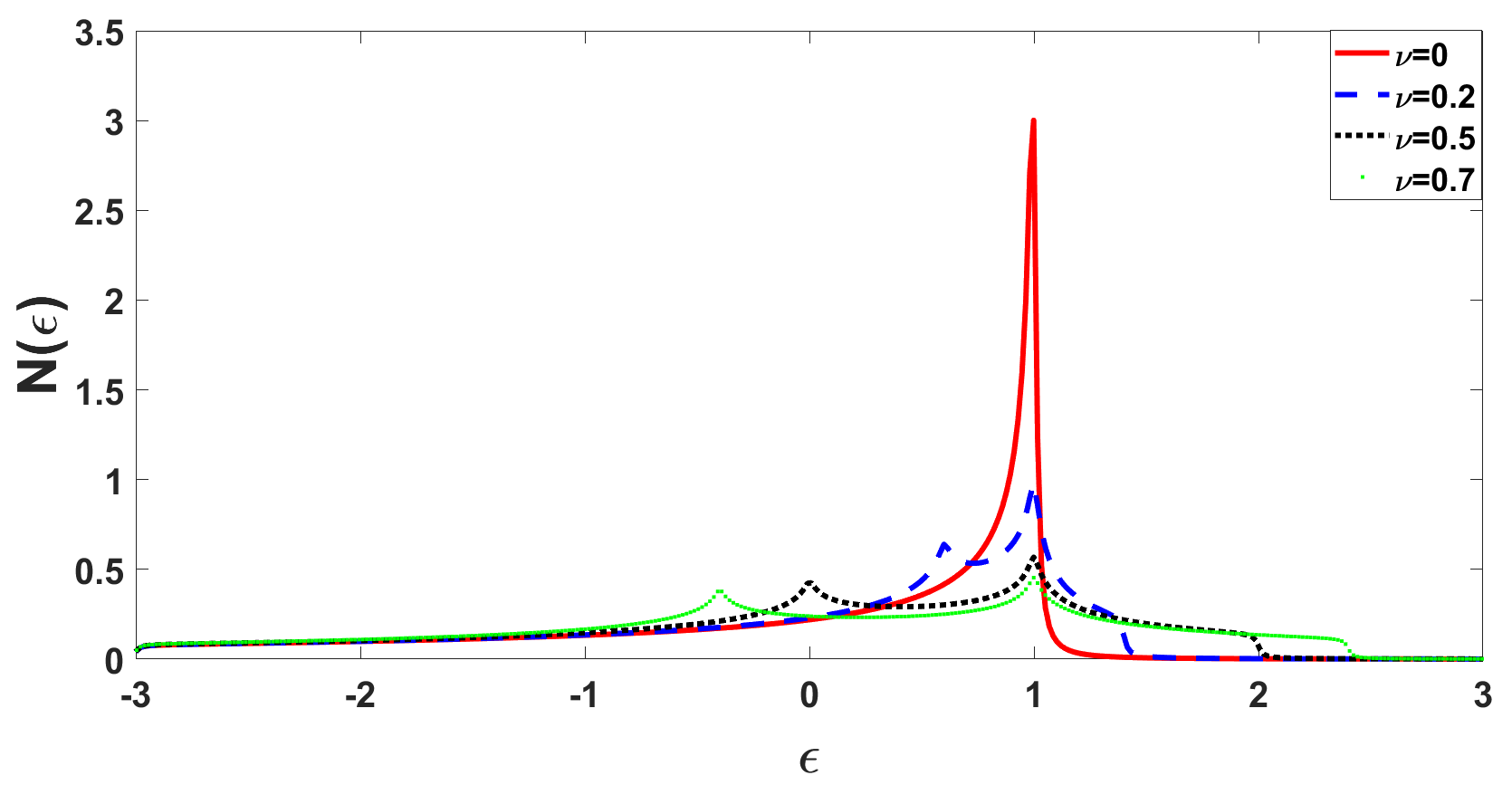}
 \caption{The DOS $N(\epsilon)$ for $\epsilon_+(\bm{k})$ given by Eq. (\ref{mh11}) with $t_1=0$ (top) and
 $\epsilon_+(\bm{k})$ given by Eq. (\ref{mh13}) with $r=0.5$ (bottom) as a function of $\epsilon$ at various 
 values of $\nu$. All energies are measured in units of $t_0$ and $N(\epsilon)$ measured in units of $1/t_0$.}
 \label{amdosf1}
\end{center}
\end{figure}

From the above analysis, whether or not the Kondo temperature is reduced depends on the value of $N(0)$. 
Figure \ref{amdosf1} shows the behavior of $N(\epsilon)$ for $\epsilon_+(\bm{k})$ given by Eq. (\ref{mh11}) 
with $t_1=0$ at various values of $\nu$. Let us understand it. The band structure has van Hove singularities 
at the points $(0,0)$, $(\pm\pi,0)$, $(0,\pm\pi)$, $(\pi,\pm\pi)$, and $(\pm\pi,\pi)$. The corresponding 
energies (in units of $t_0$) are
\begin{eqnarray*}
	-2\nu:~~(\pm\pi,0) \ , & & 2\nu:~~(0,\pm\pi) \ , \\
	-2:~~(0,0) \ , & & 2:~~\mbox{otherwise} \ .
\end{eqnarray*}
For $\nu=0$, there is a van Hove singularity at $\epsilon=0$, the band center. This van Hove singularity 
spits into a pair located at $\epsilon=\pm 2\nu$ as $\nu\neq 0$. When the Fermi energy is close to 
$\epsilon=0$ or the electron density is close to half-filling, the DOS with $\nu\neq 0$ is reduced compared 
to the one with $\nu=0$, leading to the suppression of the Kondo temperature. This is the basis for the 
claim made in Ref. \onlinecite{Diniz}. However, when the deviation of the electron density from half-filling 
is large such that the Fermi energy is far away from $\epsilon=0$, the the DOS with $\nu\neq 0$ is increased 
compared to the one with $\nu=0$, leading to the enhancement of the Kondo temperature. 

Actually, the suppression of the Kondo temperature in an AM metal as the electron density is close to 
half-filling is an artifact for the model we employed. For the generic band structure of the AM metal, the 
van Hove singularities will in general not be located at the band center. For example, let us add a 
next-nearest-neighbor term to the band structure 
\begin{eqnarray}
 \epsilon_{\bm{\sigma}}(\bm{k}) &=& -t_0(\cos{k_x}+\cos{k_y})+2\sigma t_2(\cos{k_x}-\cos{k_y}) \nonumber \\
 & & -rt_0[\cos{(k_x+k_y)}+\cos{(k_x-k_y)}] \ , \label{mh13}
\end{eqnarray}
and plot the corresponding DOS for spin-up electrons with $r=0.5$ in Fig. \ref{amdosf1}. (Note that 
the addition of this next-nearest-neighbor term does not spoil the $C_{4z}$ symmetry. Moreover, the band 
minimum is still located at the $\Gamma$ point as long as $r>-0.5$.) Now the original singularity at the 
band center is moved toward the band edge. When the electron density is near half-filling, the DOS at the 
Fermi energy is increased compared to the one with $\nu=0$ , leading to the enhancement of the Kondo 
temperature. Therefore, we conclude that whether the Kondo temperature is enhanced or suppressed in an AM 
metal depends on the band structure as well as the electron density.

%%%%%%%%%%%%%%%%%%%%%%%%%%%%%%%%%%%%%%%%%%%%%%%%%%%%%%%%%%%%%%%%%%%%%%%%%%%%%%%%%%%%%%%%%%%%%%%%%%%%%%%%%%%
\subsection{The variational wavefunction}

Now we would like to study the ground-state properties at $J_0>0$ by using the variational wavefunction 
approach. The ground state of $H_0$ is 
\begin{eqnarray*}
 |\Psi_0\rangle=\prod^{\prime}_{\bm{k},\sigma}c^{\dagger}_{\bm{k}\sigma}|0\rangle \ ,
\end{eqnarray*}
where the zero-particle state $|0\rangle$ is defined by $c_{\bm{k}\sigma}|0\rangle=0$ for all $\bm{k}$ and 
$\sigma$ and $\prod^{\prime}$ means the product over states below the Fermi energy, i.e., 
$\epsilon_{\sigma}(\bm{k})<\epsilon_F$. To study the ground-state properties of $H$, we try the following 
ansatz for the ground-state wavefunction of $H$\cite{Yosida}:
\begin{equation}
 |\Psi\rangle=\sum^{\prime}_{\bm{k},\sigma}\Gamma_{\bm{k}\sigma}c^{\dagger}_{\bm{k}-\sigma}|\Psi_0\rangle
 \otimes|\sigma\rangle \ , \label{amkondo1}
\end{equation}
where $|\sigma\rangle$ is the eigenstate of $S_z$ with eigenvalue $\sigma/2$ and $\sum^{\prime}$ means the 
sum over states above the Fermi energy, i.e., $\epsilon_{\sigma}(\bm{k})>\epsilon_F$. The coefficients 
$\Gamma_{\bm{k}\sigma}$ can be determined by requiring that the energy functional
\begin{eqnarray*}
 E=\frac{\langle\Psi|H|\Psi\rangle}{\langle\Psi|\Psi\rangle} \ ,
\end{eqnarray*}
takes the minimum value, where 
\begin{eqnarray*}
 \langle\Psi|\Psi\rangle=\sum^{\prime}_{\bm{k},\sigma}|\Gamma_{\bm{k}\sigma}|^2 \ .
\end{eqnarray*}

Straightforward calculation gives
\begin{eqnarray*}
 \langle\Psi|H|\Psi\rangle &=& \! \sum^{\prime}_{\bm{k},\sigma}|\Gamma_{\bm{k}\sigma}|^2
 [E_0+\epsilon_{-\sigma}(\bm{k})-\epsilon_F] \\
 & & +\frac{J}{V} \! \sum^{\prime}_{\bm{k},\bm{k}^{\prime},\sigma} \! 
 \left(\Gamma_{\bm{k}\sigma}^*\Gamma_{\bm{k}^{\prime}-\sigma}-\frac{1}{2}\Gamma_{\bm{k}\sigma}^*
 \Gamma_{\bm{k}^{\prime}\sigma}\right) ,
\end{eqnarray*}
where $E_0$ is the ground-state energy of $H_0$. The minimum of $E$ is determined by the equation 
$\delta E/\delta\Gamma_{\bm{k}\sigma}^*=0$, yielding
\begin{equation}
 \Gamma_{\bm{k}\sigma}=-\frac{J(\Gamma_{-\sigma}-\Gamma_{\sigma}/2)}
 {\epsilon_{-\sigma}(\bm{k})-\epsilon_F+\Delta_b} \ , \label{amkondo2}
\end{equation}
where $\Delta_b=E_0-E$ is the binding energy and
\begin{eqnarray*}
 \Gamma_{\sigma}=\frac{1}{V}\sum^{\prime}_{\bm{k}}\Gamma_{\bm{k}\sigma} \ .
\end{eqnarray*}

Equation (\ref{amkondo2}) can be written as
\begin{eqnarray*}
 \left(1-\frac{\eta}{2}\right) \! \Gamma_+=-\eta\Gamma_- \ , ~~
 \left(1-\frac{\eta}{2}\right) \! \Gamma_-=-\eta\Gamma_+ \ .
\end{eqnarray*}
provided that $\Delta_b\ll D$, where 
\begin{eqnarray*}
 \eta=N(0)J\ln{(D/\Delta_b)} \ .
\end{eqnarray*}
By requiring $\Gamma_{\pm}\neq 0$, we find that
\begin{eqnarray*}
 1-\frac{\eta}{2}=\pm\eta \ ,
\end{eqnarray*}
leading to $\eta=2/3$ and $\eta=-2$ for the upper and lower signs, respectively. Since $\eta>0$ for $J>0$, we
conclude that $\eta=2/3$ and
\begin{equation}
 \Delta_b=De^{-2/[3N(0)J]} \ , \label{amkondo14}
\end{equation}
for $N(0)J\ll 1$. Moreover, $\Gamma_-=-\Gamma_+$. This implies that the ground state is a spin singlet.
Therefore, the impurity spin is completely screened by the itinerant electrons.

A few remarks on the above results are in order. (i) First of all, it seems that we still get a solution for
$J<0$ (leading to $\eta=-2$), which corresponds to the triplet bound state. This is an artifact produced by 
this simple ansatz for the ground-state wavefunction. It will be removed if we take into account the 
particle-hole excitations in the ground-state wavefunction\cite{Yosida}. (ii) The resulting binding energy 
for $J>0$ takes the same form as the one in the ordinary FL. The only difference between an AM metal and 
the ordinary FL lies at the DOS $N(0)$. (iii) The spin-split Fermi surface in the AM metal reveals itself 
in the amplitude $\Gamma_{\bm{k}\sigma}$, which can be written as
\begin{equation}
 \Gamma_{\bm{k}\sigma}=\frac{3\sigma J\Gamma/2}{\epsilon_{-\sigma}(\bm{k})-\epsilon_F+\Delta_b} \ .
 \label{amkondo21}
\end{equation}
where $\Gamma=\Gamma_+$. This will result in an anisotropic behavior of the spin correlation function
\begin{equation}
 c(\bm{r})\equiv\langle\bm{S}\cdot\psi^{\dagger}\bm{\sigma}\psi(\bm{r})\rangle \ , \label{amkondo12}
\end{equation}
which is a measure of the extension of the singlet state. In the following, we shall calculate $c(\bm{r})$ 
at $T=0$.

%%%%%%%%%%%%%%%%%%%%%%%%%%%%%%%%%%%%%%%%%%%%%%%%%%%%%%%%%%%%%%%%%%%%%%%%%%%%%%%%%%%%%%%%%%%%%%%%%%%%%%%%%
\subsection{Spin correlations}

Before plunging into the computation of $c(\bm{r})$, we first determine $\Gamma$ by requiring that 
$\langle\Psi|\Psi\rangle=1$. Using Eq. (\ref{amkondo21}), we find that
\begin{equation}
 V|\Gamma|^2\approx\frac{2\Delta_b}{9N(0)J^2} \ , \label{amkondo13}
\end{equation}
for $N(0)J\ll 1$.

At $T=0$, $c(\bm{r})$ can be written as
\begin{eqnarray*}
 c(\bm{r}) &=& -\frac{\Delta_b}{2N(0)}I_-^*(\bm{r})I_+(\bm{r})+\mathrm{C.c.} \\
 & & -\frac{\Delta_b}{4N(0)}\sum_{\sigma}|I_{\sigma}(\bm{r})|^2 \ ,
\end{eqnarray*}
where we have used Eq. (\ref{amkondo13}) to eliminate $J^2V|\Gamma|^2$ and 
\begin{eqnarray*}
 I_{\sigma}(\bm{r})=\frac{1}{V} \! \sum^{\prime}_{\bm{k}}\frac{e^{i\bm{k}\cdot\bm{r}}}
 {\epsilon_{\sigma}(\bm{k})-\epsilon_F+\Delta_b} \ .
\end{eqnarray*}
Instead of an exact evaluation of $I_{\sigma}(\bm{r})$, we will consider its behaviors at $k_Fr\ll 1$ and
$k_Fr\gg 1$, where $k_F=\sqrt{2m\epsilon_F}$.

\begin{figure}
	\begin{center}
		\includegraphics[width=0.99\columnwidth]{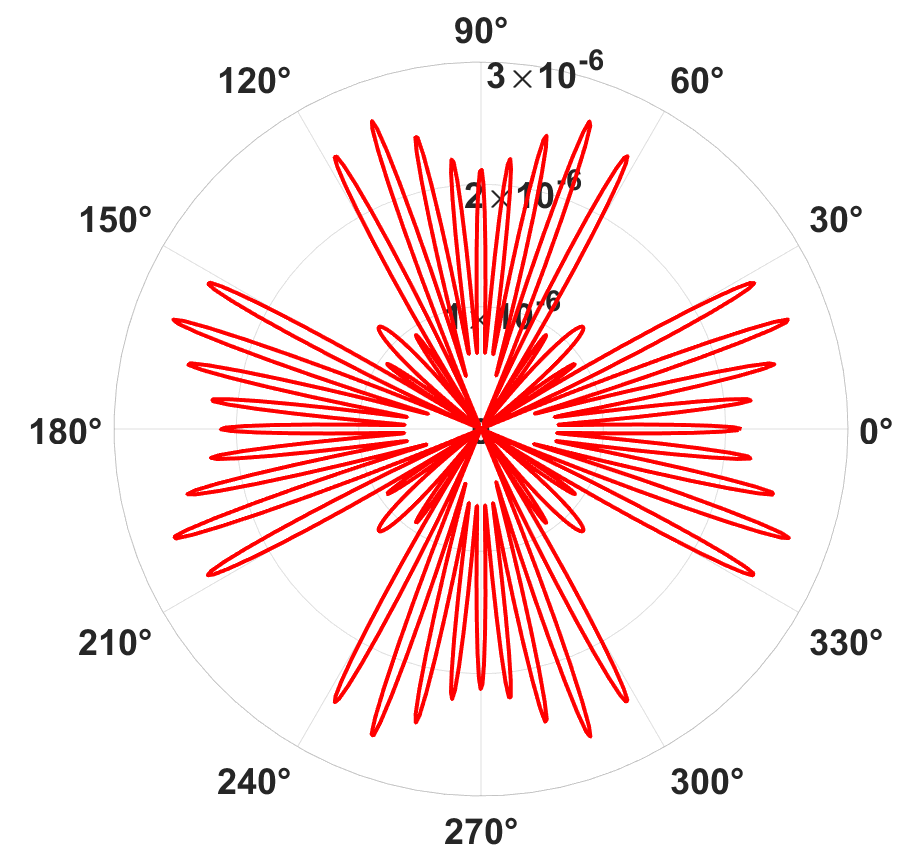}
		\caption{Angular dependence of $|c(\bm{r})|$ with $k_F\xi=10=r/\xi$. We take $\nu=0.3$ and $\phi=0$ (or 
			$\alpha_1=0.3$ and $\alpha_2=0$). $|c(\bm{r})|$ is measured in units of $N(0)\epsilon^2_F/(\pi^3\Delta_b)$.}
		\label{amkondof1}
	\end{center}
\end{figure}

The details of the calculation are left in App. \ref{a1}. Here we just list the results. For $k_Fr\ll 1$,
\begin{equation}
 I_{\sigma}(\bm{r})\approx\frac{1}{3\pi J} \! \left[1-\frac{(L_{\sigma}k_Fr)^2}{4}\right] \! -\frac{N(0)}
 {8\pi}(k_DL_{\sigma}r)^2 \ . \label{amkondo15}
\end{equation}
Since $N(0)J\ll 1$, the last term in Eq. (\ref{amkondo15}) can be neglected. On the other hand,
\begin{eqnarray}
 I_{\sigma}(\bm{r}) \! \! &=& \! \! -\sqrt{\frac{2}{\pi}}\frac{N(0)\epsilon_F}{\pi\Delta_b}
 \frac{\sin{(L_{\sigma}k_Fr-\pi/4)}}{(L_{\sigma}k_Fr)^{3/2}} \nonumber \\
 \! \! & & \! \! +\sqrt{\frac{2}{\pi}}\frac{N(0)\epsilon_F}{\pi\Delta_b} \! \left(\frac{2\epsilon_F}
 {\Delta_b}-\frac{1}{8}\right) \! \frac{\cos{(L_{\sigma}k_Fr-\pi/4)}}{(L_{\sigma}k_Fr)^{5/2}} \nonumber \\
 & & +O \! \left[(L_{\sigma}k_Fr)^{-7/2}\right] , \label{amkondo16}
\end{eqnarray}
for $k_Fr\gg 1$. For $r\gg\xi$, the second term is much smaller than the first one and can be neglected,
where $\xi=2\epsilon_F/(k_F\Delta_b)$ is the Kondo screening length. Since $\epsilon_F/\Delta_b\gg 1$, 
$\xi\gg1/k_F$. In the above, 
\begin{eqnarray*}
 L_{\sigma}=\sqrt{1+\sigma\nu\sin{[2(\theta-\phi)]}} \ ,
\end{eqnarray*}
with $x=r\cos{\theta}$ and $y=r\sin{\theta}$. Note that $L_{\sigma}(-\hat{\bm{r}})=L_{\sigma}(\hat{\bm{r}})$ 
where $\hat{\bm{r}}=\bm{r}/r$. Moreover, $L_{\sigma}(\bm{r})$ transform as
\begin{equation}
 L_{\sigma}(C_{4z}\hat{\bm{r}})=L_{-\sigma}(\hat{\bm{r}}) \ , \label{amkondo22}
\end{equation}
under the $C_{4z}$ transformation.

\begin{figure}
	\begin{center}
		\includegraphics[width=0.99\columnwidth]{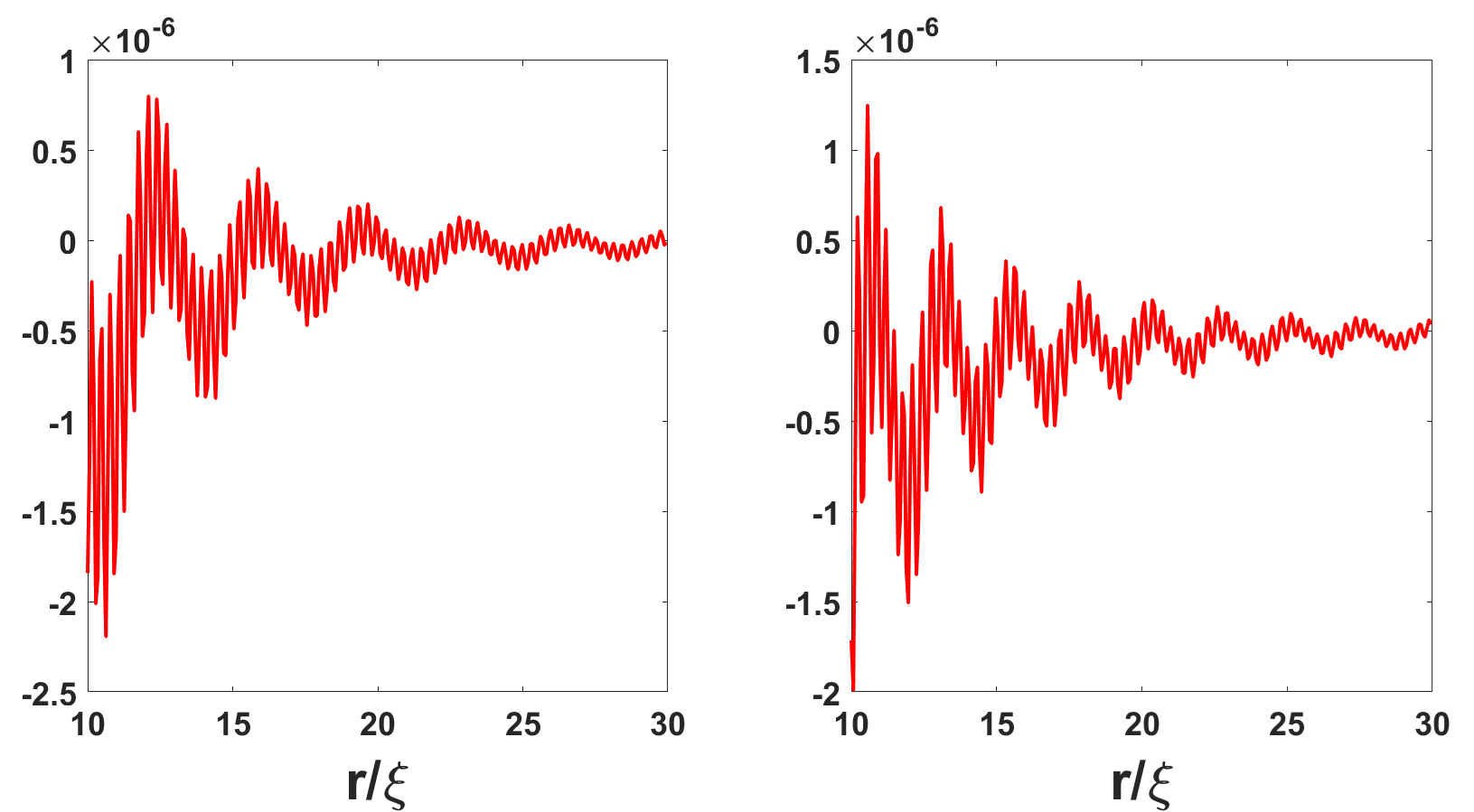}
		\caption{Radial dependence of $c(\bm{r})$ at $r/\xi\gg 1$ for $\theta=\pi/10$ (left) and $\pi/3$ (right), with 
			$k_F\xi=10$. We take $\nu=0.3$ and $\phi=0$ (or $\alpha_1=0.3$ and $\alpha_2=0$). $c(\bm{r})$ is measured in 
			units of $N(0)\epsilon^2_F/(\pi^3\Delta_b)$.}
		\label{amkondof11}
	\end{center}
\end{figure}

In terms of Eqs. (\ref{amkondo15}) and (\ref{amkondo16}), we find that 
\begin{equation}
 c(\bm{r})\approx -\frac{5\Delta_b}{24\pi^2N(0)J^2} \! \left[1-\frac{3}{5}(k_Fr)^2\right] , \label{amkondo17}
\end{equation}
for $k_Fr\ll 1$, and
\begin{widetext}
\begin{equation}
 c(\bm{r})\approx -\frac{N(0)\epsilon_F^2}{\pi^3\Delta_b} \! \left[
 \frac{\cos{[(L_+-L_-)k_Fr]}-\sin{[(L_++L_-)k_Fr]}}{(k_Fr)^3}+\frac{1}{2}\sum_{\sigma}
 \frac{\sin^2{(L_{\sigma}k_Fr)}}{(L_{\sigma}k_Fr)^3}\right] , \label{amkondo18}
\end{equation}
\end{widetext}
for $r\gg\xi$. A few remarks on the above results are in order. (i) First of all, the $d$-wave nature 
of the Fermi surface is reflected in the long distance behavior of $c(\bm{r})$. That is, 
$c(C_{4z}\bm{r})=c(\bm{r})$ as illustrated in Fig. \ref{amkondof1}. This results from the interference 
between the spin-up and -down Fermi surfaces, as shown in the first term in the bracket in Eq. 
(\ref{amkondo18}).  In contrast, this interference disappears at the short distance such that $c(\bm{r})$ 
is almost isotropic, as can be seen from Eq. (\ref{amkondo17}) which is independent of $\theta$. (ii) 
Next, $c(\bm{r})$ exhibits an oscillatory behavior with the amplitude decaying as $1/r^3$ at long 
distances. The oscillation in general has four periods 
\begin{eqnarray*}
 \frac{2\pi}{(L_+-L_-)k_F} \ , ~\frac{2\pi}{(L_++L_-)k_F} \ ,~\frac{\pi}{L_{\pm}k_F} \ ,
\end{eqnarray*}
which all depend on the direction of observation, as illustrated in Fig. \ref{amkondof11}. In the limit 
$\nu\rightarrow 0$, these periods reduce to a single one $\pi/k_F$, identical to that in an ordinary FL\cite{Ishii}.
Thus, this multi-period oscillation of the Kondo screening cloud also follows from the interference 
between the spin-up and -down Fermi surfaces.

%%%%%%%%%%%%%%%%%%%%%%%%%%%%%%%%%%%%%%%%%%%%%%%%%%%%%%%%%%%%%%%%%%%%%%%%%%%%%%%%%%%%%%%%%%%%%%%%%%%%%%%%%
\section{The RKKY interaction}
\label{rkky}

Now we consider the RKKY interaction between two magnetic impurities induced by the AM metal. The exchange 
interaction between the magnetic impurities and itinerant electrons can be written as
\begin{equation}
 H_{ex}=J\sum_{i=1,2}\bm{S}_i\cdot\psi^{\dagger}\bm{\sigma}\psi(\bm{r}_i) \ , \label{amh14}
\end{equation}
where $\bm{S}_i$ is the spin of the magnetic impurity at the position $\bm{r}_i$. Our working Hamiltonian
is $H=H_0+H_{ex}$. We further assume that the separation between the two spins $|\bm{r}_1-\bm{r}_2|$ is 
much smaller than the size of the Kondo screening cloud, i.e., $|\bm{r}_1-\bm{r}_2|\ll\xi$, so that the 
Kondo effect can be neglected.

The RKKY interaction is given by
\begin{equation}
 V_{ab}(\bm{r}_1-\bm{r}_2)=J^2\Pi_{ab}(\bm{r}_1-\bm{r}_2) \ , \label{amrkky1}
\end{equation}
where in the imaginary-time formulation, 
\begin{equation}
 \Pi_{ab}(\bm{r}_1-\bm{r}_2)=\! \int^{\beta}_0 \! d\tau\mathcal{S}_{ab}(\tau,\bm{r}_1-\bm{r}_2) \ , 
 \label{amrkky11}
\end{equation}
and
\begin{equation}
 \mathcal{S}_{ab}(X_1-X_2)\equiv-\langle\mathcal{T}_{\tau}\{\hat{O}_a(X_1)\hat{O}_b(X_2)\}\rangle \ , 
 \label{amrkky12}
\end{equation}
is the spin-spin correlation function for the AM metal. In Eq. (\ref{amrkky12}), $X=(\tau,\bm{r})$,
$\hat{O}_a=\psi^{\dagger}\sigma_a\psi$, and $\mathcal{T}_{\tau}$ denotes the time ordering.

For non-interacting fermions, $\mathcal{S}_{ab}(X_1-X_2)$ can be written as
\begin{eqnarray*}
 \mathcal{S}_{ab}(X_1-X_2)=\Lambda^{ab}_{\alpha\beta\lambda\rho}G^{(0)}_{\beta\lambda}(X_1-X_2)
 G^{(0)}_{\rho\alpha}(X_2-X_1) \ ,
\end{eqnarray*}
where $\Lambda^{ab}_{\alpha\beta\lambda\rho}=(\sigma_a)_{\alpha\beta}(\sigma_b)_{\lambda\rho}$ and 
\begin{eqnarray*}
 G^{(0)}_{\alpha\beta}(X_1-X_2)=-\langle\mathcal{T}_{\tau}\{\psi_{\alpha}(X_1)\psi^{\dagger}_{\beta}(X_2)\}
 \rangle_0 \ , 
\end{eqnarray*}
is the single-particle Green function for non-interacting fermions, with $\langle\cdots\rangle_0$ denotes 
the average with respect to $H_0$. For the AM metal described by $H_0$, 
\begin{eqnarray*}
 G^{(0)}_{\alpha\beta}(X_1-X_2)=\delta_{\alpha\beta}G_{\alpha}^{(0)}(X_1-X_2) \ ,
\end{eqnarray*}
and thus we have
\begin{eqnarray*}
 \mathcal{S}_{ab}(X_1-X_2) \! = \! \Lambda^{ab}_{\alpha\beta\beta\alpha}G^{(0)}_{\beta}(X_1-X_2)G^{(0)}_{\alpha}
 (X_2-X_1) \ .
\end{eqnarray*}
Let $\mathcal{G}_{\alpha}^{(0)}(i\omega_n,\bm{r})$ be the Fourier transform of $G^{(0)}_{\alpha}(\tau,\bm{r})$,
i.e.,
\begin{eqnarray*}
 G^{(0)}_{\alpha}(\tau,\bm{r})=\frac{1}{\beta}\sum_ne^{-i\omega_n\tau}\mathcal{G}_{\alpha}^{(0)}
 (i\omega_n,\bm{r}) \ ,
\end{eqnarray*}
where $\omega_n=(2n+1)\pi T$. Then, $\Pi_{ab}$ becomes
\begin{eqnarray*}
 \Pi_{ab}(\bm{r})=\frac{1}{\beta}\Lambda^{ab}_{\alpha\beta\beta\alpha} \! \sum_n\mathcal{G}_{\beta}^{(0)}
 (i\omega_n,\bm{r})\mathcal{G}_{\alpha}^{(0)}(i\omega_n,-\bm{r}) \ .
\end{eqnarray*} 

To proceed, we insert the spectral representation of $\mathcal{G}_{\alpha}^{(0)}(i\omega_n,\bm{r})$
\begin{eqnarray*}
 \mathcal{G}_{\alpha}^{(0)}(i\omega_n,\bm{r})=\! \int^{+\infty}_{-\infty} \! \frac{d\nu}{2\pi}
 \frac{\rho_{\alpha}(\nu,\bm{r})}{i\omega_n-\nu} \ ,
\end{eqnarray*}
into the above relation, and find that
\begin{eqnarray*}
 \Pi_{ab}(\bm{r}) &=& \Lambda^{ab}_{\alpha\beta\beta\alpha} \! \int^{+\infty}_{-\infty} \! 
 \frac{d\omega_1d\omega_2}{(2\pi)^2}\rho_{\beta}(\omega_1,\bm{r})\rho_{\alpha}(\omega_2,-\bm{r}) \\
 & & \times\frac{n_F(\omega_1)-n_F(\omega_2)}{\omega_1-\omega_2} \ ,
\end{eqnarray*}
where $n_F(x)=1/(e^{\beta x}+1)$. At $T=0$, $n_F(\omega)=\Theta(-\omega)$ and $\Pi_{ab}(\bm{r})$ reduces to
\begin{eqnarray*}
 \Pi_{ab}(\bm{r})=W_{ab}(\bm{r})+W_{ba}(-\bm{r}) \ ,
\end{eqnarray*}
where
\begin{eqnarray*}
 W_{ab}(\bm{r}) &=& \Lambda^{ab}_{\alpha\beta\beta\alpha} \! \int^{+\infty}_{-\infty} \! 
 \frac{d\omega_1d\omega_2}{(2\pi)^2}\rho_{\beta}(\omega_1,\bm{r})\rho_{\alpha}(\omega_2,-\bm{r}) \\
 & & \times\frac{\Theta(-\omega_1)}{\omega_1-\omega_2} \ .
\end{eqnarray*}
In the above, we have used the identity
\begin{eqnarray*} 
 \Lambda^{ba}_{\alpha\beta\lambda\rho}=\Lambda^{ab}_{\lambda\rho\alpha\beta} \ .
\end{eqnarray*} 
Note that $\Pi_{ba}(\bm{r})=\Pi_{ab}(-\bm{r})$. 

The rest of calculations is left to App. \ref{a2}. Here we just list the results. The only nonvanishing 
components of $\Pi_{ab}(\bm{r})$ are 
\begin{widetext}
\begin{equation}
 \Pi_{33}(\bm{r})=\frac{mk_F^2}{4\pi(1-\nu^2)} \! \sum_{\sigma}
 [J_0(k_FrL_{\sigma})Y_0(k_FrL_{\sigma})+J_1(k_FrL_{\sigma})Y_1(k_FrL_{\sigma})] \ , \label{amrkky13}
\end{equation}
and
\begin{equation}
 \Pi_{\perp}(\bm{r})=\frac{mk_F^2}{\pi(1-\nu^2)}
 \frac{L_+J_1(k_FrL_+)Y_0(k_FrL_-)-L_-J_1(k_FrL_-)Y_0(k_FrL_+)}{k_Fr(L_+^2-L_-^2)} \ , \label{amrkky14}
\end{equation}
\end{widetext}
where $\Pi_{11}(\bm{r})=\Pi_{22}(\bm{r})\equiv\Pi_{\perp}(\bm{r})$, $J_{\nu}(x)$ is the Bessel function 
of the first kind, and $Y_{\nu}(x)$ is the Bessel function of the second kind. The appearance of the 
Bessel functions is quite common in the RKKY interaction in $d=2$.\cite{Dugaev,Saremi,Kogan,Liu} Therefore, 
the interaction between two local spins is of the form
\begin{eqnarray}
 H_{12} &=& J^2\Pi_{\perp}(\bm{r}_1-\bm{r}_2)(S_1^xS_2^x+S_1^yS_2^y) \nonumber \\
 & & +J^2\Pi_{33}(\bm{r}_1-\bm{r}_2)S_1^zS_2^z \ . \label{amrkky15}
\end{eqnarray}
We see that the RKKY interaction produced by an AM metal is of the anisotropic Heisenberg type. This 
results from the spin splitting Fermi surfaces, which can be seen by noting that 
$\Pi_{33}(\bm{r})=\Pi_{\perp}(\bm{r})$ when we set $t_1=0=t_2$. Since $L_+\leftrightarrow L_-$ under the 
$C_{4z}$ rotation, we find that both $\Pi_{\perp}(\bm{r})$ and $\Pi_{33}(\bm{r})$ preserve this symmetry. 
That is,
\begin{eqnarray*}
 \Pi_{\perp}(C_{4z}\bm{r})=\Pi_{\perp}(\bm{r}) \ , ~~\Pi_{33}(C_{4z}\bm{r})=\Pi_{33}(\bm{r}) \ .
\end{eqnarray*}
Moreover, both $\Pi_{\perp}(\bm{r})$ and $\Pi_{33}(\bm{r})$ are invariant under space inversion because 
$L_{\sigma}(-\hat{\bm{r}})=L_{\sigma}(\hat{\bm{r}})$.

\begin{figure}
	\begin{center}
		\includegraphics[width=0.99\columnwidth]{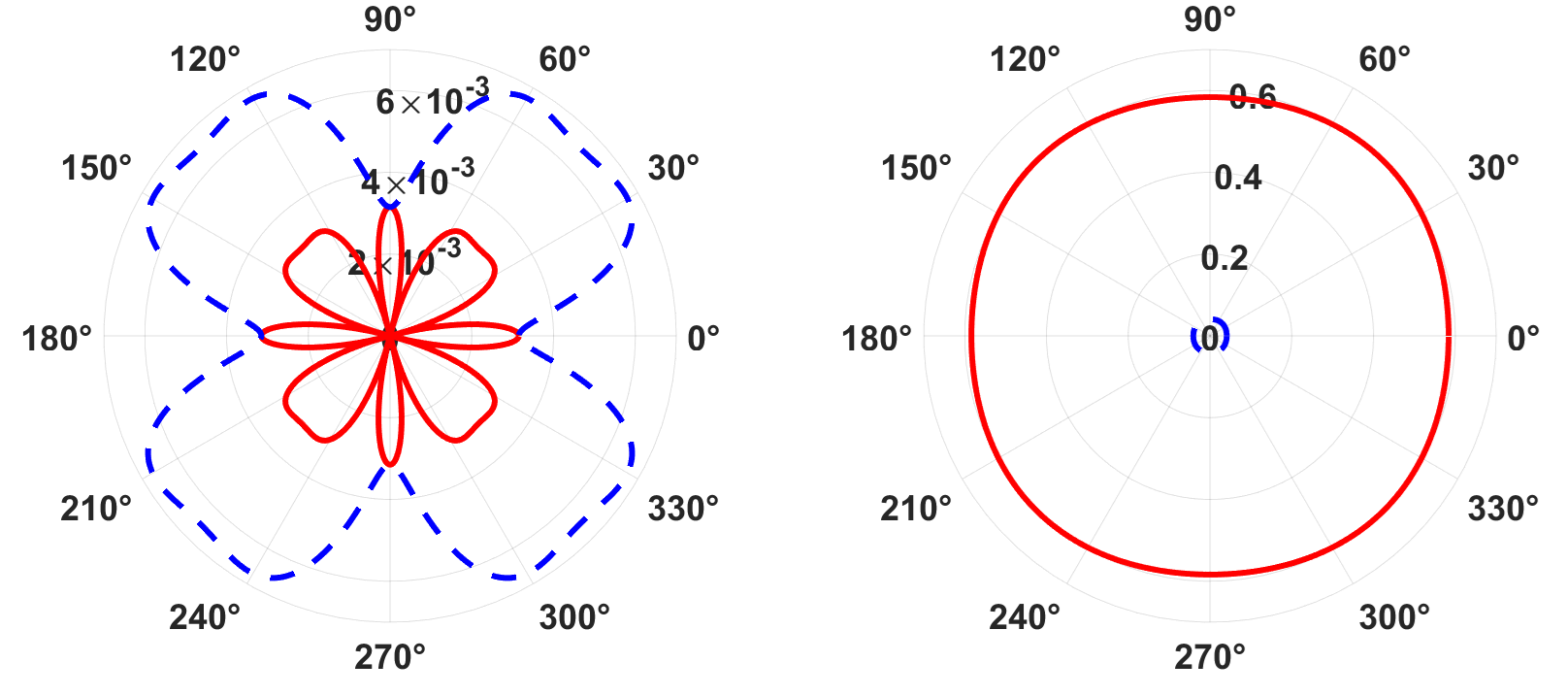}
		\caption{Angular dependence of $|\Pi_{\perp}(\bm{r})|$ (dashed lines) and $|\Pi_{33}(\bm{r})|$ (solid lines)
			with $k_Fr=10$ (left) and $k_Fr=0.7$ (right). We take $\nu=0.3$ and $\phi=0$ (or $\alpha_1=0.3$ and 
			$\alpha_2=0$). Both are measured in units of $mk_F^2/(2\pi)$.}
		\label{amrkkyf1}
	\end{center}
\end{figure}

Let us examine the asymptotic behaviors of $\Pi_{\perp}(\bm{r})$ and $\Pi_{33}(\bm{r})$. Using the 
asymptotic behaviors of the Bessel functions at the short and long distances\cite{Grad}, we find that
\begin{equation}
 \Pi_{33}(\bm{r})\sim\frac{mk_F^2}{\pi^2(1-\nu^2)} \! \left[\ln{\! \left(\frac{k_F^2r^2L_+L_-}{4}\right)} 
 \! +2\gamma-\frac{1}{2}\right] , \label{amrkky16}
\end{equation}
and
\begin{eqnarray}
 \Pi_{\perp}(\bm{r}) &\sim& \frac{mk_F^2}{2\pi^2(1-\nu^2)} \! \left[\ln{\! \left(\frac{k_F^2r^2L_+L_-}{4}\right)} 
 \right. \nonumber \\
 & & -\! \left.\frac{\ln{(L_+/L_-})}{\nu\sin{[2(\theta-\phi)]}}+\gamma\right] , \label{amrkky17}
\end{eqnarray}
as $k_Fr\ll 1$ and
\begin{equation}
 \Pi_{33}(\bm{r})\sim-\frac{mk_F^2}{\pi^2(1-\nu^2)}\sum_{\sigma}
 \frac{\sin{(2k_FrL_{\sigma})}}{(2k_FrL_{\sigma})^2} \ , \label{amrkky18}
\end{equation}
and
\begin{eqnarray}
 \Pi_{\perp}(\bm{r}) \! \! &\sim& \! \! \frac{mk_F^2}{\pi^2(1-\nu^2)} \nonumber \\
 \! \! & & \! \! \times \! \left[\frac{\cos{[k_Fr(L_+-L_-)]} \! - \! \sin{[k_Fr(L_++L_-)]}}
 {k_F^2r^2(L_++L_-)\sqrt{L_+L_-}}\right] , \nonumber \\
 \! \! & & \! \! \label{amrkky19}
\end{eqnarray}
as $k_Fr\gg 1$, where $\gamma=0.577215\cdots$ is the Euler constant.

Similar to the FL in $d=2$, both $\Pi_{\perp}(\bm{r})$ and $\Pi_{33}(\bm{r})$ decay as $1/r^2$ at long 
distances\cite{Kittel}. However, in the present case, they exhibit nontrivial angular dependence due to 
the underlying spin splitting Fermi surfaces. Moreover, this directional dependence reflects that of the 
Fermi surfaces. That is, they exhibit the $C_{4z}$ symmetry. In contrast, at short distances, i.e., 
$k_Fr\ll 1$, both are nearly isotropic, similar to the case in an ordinary FL. Moreover, the magnitude of 
$\Pi_{\perp}(\bm{r})$ is much smaller than that of $\Pi_{33}(\bm{r})$. This results in an Ising type of 
exchange interaction between the local spins. However, as $k_Fr\gg 1$, the situation is reversed: the 
magnitude of $\Pi_{\perp}(\bm{r})$ becomes larger than that of $\Pi_{33}(\bm{r})$, leading to the $XY$ 
type of exchange interaction between the local spins. We plot the angular dependence of $\Pi_{\perp}(\bm{r})$ 
and $\Pi_{33}(\bm{r})$ in Fig. \ref{amrkkyf1}. 

\begin{figure}
	\begin{center}
		\includegraphics[width=0.99\columnwidth]{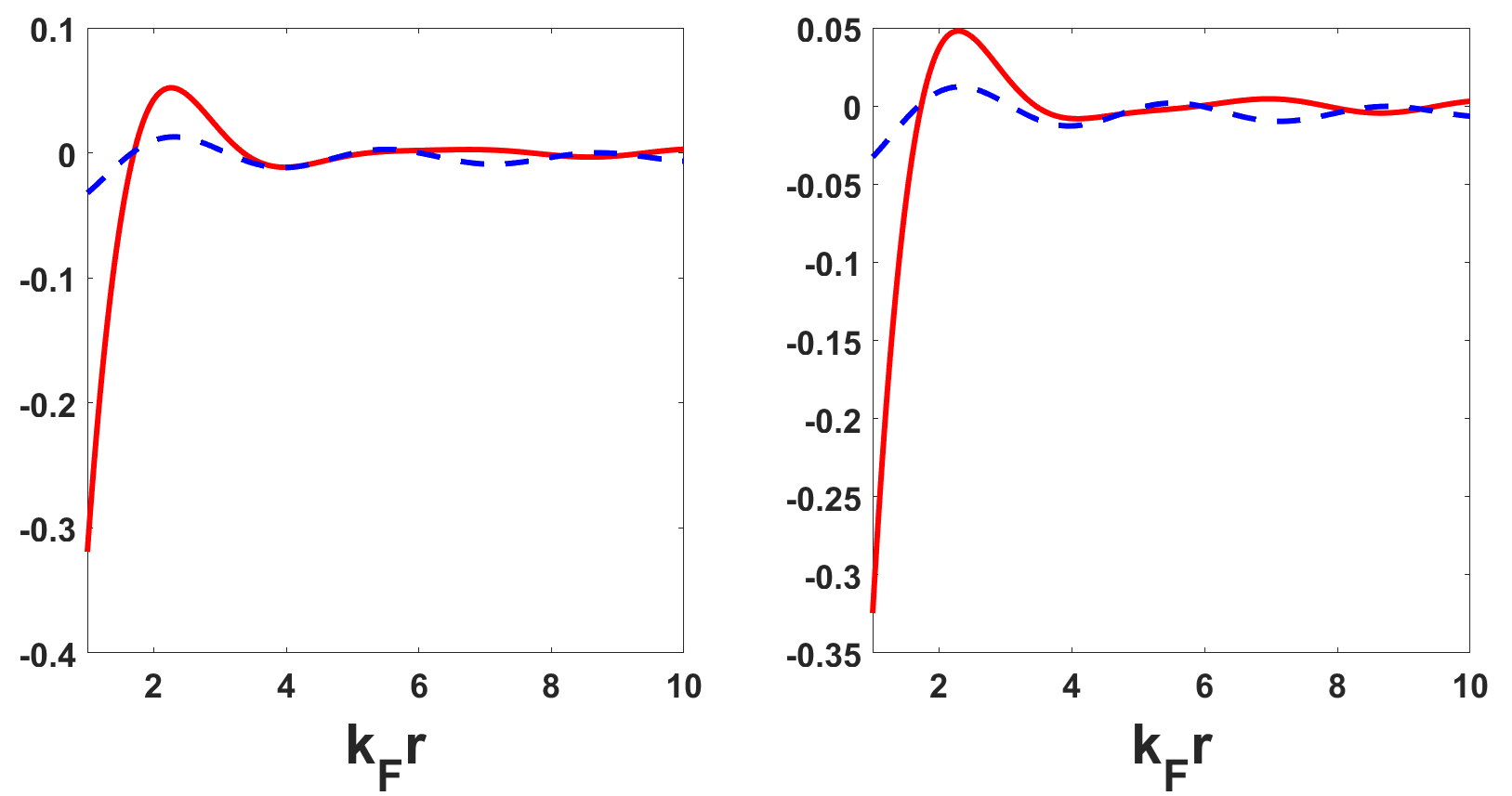}
		\caption{The $r$ dependence of $\Pi_{\perp}(\bm{r})$ (dashed lines) and $\Pi_{33}(\bm{r})$ (solid lines)
			starting from $k_Fr=1$, with $\theta=\pi/6$ (left) and $\theta=\pi/4$ (right). We take $\nu=0.3$ and 
			$\phi=0$ (or $\alpha_1=0.3$ and $\alpha_2=0$). Both are measured in units of $mk_F^2/(2\pi)$.}
		\label{amrkkyf11}
	\end{center}
\end{figure}

We also plot the $r$ dependence of $\Pi_{\perp}(\bm{r})$ and $\Pi_{33}(\bm{r})$ in Fig. \ref{amrkkyf11},
with two different directions. As we have expected, both exhibit oscillatory power-law decay at long 
distance. This behavior is insensitive to the direction. It is the periods of oscillations which depend 
on the direction. For $\Pi_{\perp}(\bm{r})$, there are two periods of oscillations:
\begin{eqnarray*}
 \frac{2\pi}{(L_+-L_-)k_F} \ , ~~\frac{2\pi}{(L_++L_-)k_F} \ .
\end{eqnarray*}
For $\Pi_{33}(\bm{r})$, there are also two periods of oscillations: $\pi/(L_{\pm}k_F)$, which are 
distinct to those in $\Pi_{\perp}(\bm{r})$. In the limit $\nu\rightarrow 0$, all reduces to a single one
period $\pi/k_F$, as that in the ordinary FL\cite{Kittel}.

%%%%%%%%%%%%%%%%%%%%%%%%%%%%%%%%%%%%%%%%%%%%%%%%%%%%%%%%%%%%%%%%%%%%%%%%%%%%%%%%%%%%%%%%%%%%%%%%%%%%%%%%%%%
\section{Conclusions and discussions}

In the present work, we explore the effects of the spin splitting Fermi surfaces in the AM metal on the 
magnetic impurities. For the single-impurity case, though the T symmetry is broken in the host metal, we 
show that the impurity spin is still completely screened at low energies in the AM metal when the coupling 
between the local spin and itinerant electrons is AF by combining the one-loop RG equation and the 
variational wavefunction. The complete screening of the impurity spin at $T=0$ was also justified by a 
recent study on the single-impurity Anderson model based on the numerical RG\cite{Diniz}. In contrast to 
the conclusion in Ref. \onlinecite{Diniz}, we indicates that the the Kondo temperature in an AM metal may 
be enhanced or reduced, depending on the band structure and the electron density.

We further use the variational wavefunction to calculate the spin correlation at $T=0$, which measures the 
spatial extension of the Kondo screening cloud. We find that the interference between the spin-up and -down 
Fermi surfaces in the AM metal is reflected in the angular dependence of the spin correlation as well as 
the multi-periods of oscillations in its radial dependence. This is the main distinction between the Kondo 
effects in the AM metal and the ordinary FL.

The local spin is only partially screened in an ordinary AF metal and the unscreened moment fraction depends 
on the amplitude of the AF order though the AF exchange interaction between the impurity spin and conduction 
electrons also flows to the strong coupling regime\cite{AVV}. This follows from the spin nonconserving 
interaction vertices generated by the presence of the AF order. For the simple model describing the AM metal 
[$\mathcal{H}(\bm{k})$ in Eq. (\ref{mh1}) or $H_0$ in Eq. (\ref{amh1})], no such spin nonconserving 
interaction vertices are generated when the AM order is built up. In addition, the Kondo screening of an 
impurity spin in an ordinary AF metal depends on its location within the unit cell due to the nonvanishing 
ordering wavevector $\bm{Q}$ in it. However, this is not the case in an AM metal as long as Eq. (\ref{amh12}) 
is satisfied. Once Eq. (\ref{amh12}) is violated, the band structure will have global minima away from the 
$\Gamma$ point. In that case, we expect that similar phenomena may also occur in the AM metal.

The spin splitting Fermi surfaces in the AM metal also manifest themselves in the RKKY interaction between 
two magnetic impurities. We show that the RKKY interaction is of an anisotropic Heisenberg type, distinct 
from the one in an ordinary FL, which also arises from the interference between the spin-up and -down 
Fermi surfaces. The RKKY interaction in an AM metal shows the underlying $C_{4z}$ symmetry. Both the 
transverse and longitudinal components also exhibit distinct periods of oscillations, a consequence of the 
interference effect. Moreover, they have different magnitudes such that the resulting exchange interaction 
between magnetic impurities is of the Ising type at short distances and of the $XY$ type at long distances. 
Thus, one may tune the type of exchange interaction by varying the distance between magnetic impurities. 

At long distances, the RKKY interaction induced by a $2$D AM metal decays as $1/r^2$, identical to that 
produced by a FL in $d=2$. This is not the case in some $2$D systems like graphene\cite{Dugaev,Saremi,Kogan} 
or the surface of a topological insulator\cite{Liu}. In these cases, the RKKY interaction decays as $1/r^3$
due to the linear dispersion of the quasiparticles near the Fermi level. 

Finally, in the present work, we employ a continuum model [Eq. (\ref{amh1})] to calculate the RKKY 
interaction, which allows an analytic calculation. However, this is valid only at long distances, i.e., the 
length scale much larger than the Fermi wavelength. To obtain a result valid on all length scales, one must 
use a lattice Hamiltonian like Eq. (\ref{mh1}). For any bipartite lattice at half-filling with hopping 
only between AB sublattices, such as the undoped graphene, it was shown that the RKKY interaction is AF 
between impurities on opposite sublattices and is FM between impurities on the same sublattices\cite{Saremi}. 
Could similar results be drawn for an AM metal since the altermagnet also consists of two sublattices? Or 
does the RKKY interaction exhibit distinct features when the two magnetic impurities are placed on opposite 
sublattices and the same sublattices? This is an open problem and beyond the scope of the present work. 

%%%%%%%%%%%%%%%%%%%%%%%%%%%%%%%%%%%%%%%%%%%%%%%%%%%%%%%%%%%%%%%%%%%%%%%%%%%%%%%%%%%%%%%%%%%%%%%%%%%%%%%%%%%
%\backmatter
%\bmhead{Acknowledgements}
\acknowledgments

The author is grateful to the insightful discussions with M.F. Yang and Y.-W. Lee.

%%%%%%%%%%%%%%%%%%%%%%%%%%%%%%%%%%%%%%%%%%%%%%%%%%%%%%%%%%%%%%%%%%%%%%%%%%%%%%%%%%%%%%%%%%%%%%%%%%%%%%%%%%%
%\begin{appendices}
\appendix
\section{Derivation of Eqs. (\ref{amkondo15}) and (\ref{amkondo16})}
\label{a1}

To calculate the momentum integral in $I_{\sigma}(\bm{r})$, we write $\epsilon_{\bm{k}\sigma}$ as
\begin{eqnarray*}
 \epsilon_{\sigma}(\bm{k})=\frac{1}{2m}K^tM_{\sigma}K \ ,
\end{eqnarray*}
where $K=(k_x,k_y)^t$ and
\begin{eqnarray*}
 M_{\sigma}=\! \left(\begin{array}{cc}
		1-2\sigma\alpha_2 & \sigma\alpha_1 \\
		\sigma\alpha_1 & 1+2\sigma\alpha_2
 \end{array}\right) .
\end{eqnarray*}
$M_{\sigma}$ is a real symmetric matrix, which can be diagonalized by the orthogonal matrix
\begin{eqnarray*}
 U_{\sigma}=\! \left(\begin{array}{cc}
		\sigma\alpha_1/N_{\sigma+} & \sigma\alpha_1/N_{\sigma-} \\
		& \\
		\frac{\lambda_+-1+2\sigma\alpha_2}{N_{\sigma+}} & \frac{\lambda_--1+2\sigma\alpha_2}{N_{\sigma-}}
 \end{array}\right) ,
\end{eqnarray*}
such that
\begin{eqnarray*}
 U_{\sigma}^tM_{\sigma}U_{\sigma}=D=\mbox{diag}[\lambda_+,\lambda_-] \ ,
\end{eqnarray*}
where 
\begin{eqnarray*}
 \lambda_{\pm}=1\pm\sqrt{\alpha_1^2+4\alpha_2^2}=1\pm\nu \ ,
\end{eqnarray*}
are the eigenvalues of $M_{\sigma}$ and 
\begin{eqnarray*}
 N_{\sigma\pm} &=& \sqrt{2(\lambda_{\pm}-1)(\lambda_{\pm}-1+2\sigma\alpha_2)} \\
 &=& \nu\sqrt{2[1\pm\sigma\sin{(2\phi)}]} \ .
\end{eqnarray*}

By defining 
\begin{eqnarray*}
 \tilde{K}=U_{\sigma}^tK=(\tilde{k}_x,\tilde{k}_y)^t \ ,
\end{eqnarray*}
$\epsilon_{\sigma}(\bm{k})$ becomes
\begin{eqnarray*}
 \epsilon_{\sigma}(\bm{k})=\frac{1}{2m}\tilde{K}^tD\tilde{K}=\frac{\lambda_+}{2m}\tilde{k}_x^2
 +\frac{\lambda_-}{2m}\tilde{k}_y^2 \ .
\end{eqnarray*}
Thus, $I_{\sigma}(\bm{r})$ can be written as
\begin{eqnarray*}
 I_{\sigma}(\bm{r})=\frac{2m}{\sqrt{1-\nu^2}} \! \int^{\prime} \! \frac{d^2k}{(2\pi)^2}
 \frac{e^{i\bm{k}\cdot\bm{R}_{\sigma}}}{\bm{k}^2-k_F^2+k_b^2} \ ,
\end{eqnarray*}
where $\bm{R}_{\sigma}=(\tilde{x}_{\sigma}/\sqrt{\lambda_+},\tilde{y}_{\sigma}/\sqrt{\lambda_-})$ with
\begin{eqnarray*}
 \tilde{x}_{\sigma} &=& \frac{\sigma\alpha_1x+(\lambda_+-1+2\sigma\alpha_2)y}{N_{\sigma+}} \ , \\
 \tilde{y}_{\sigma} &=& \frac{\sigma\alpha_1x+(\lambda_--1+2\sigma\alpha_2)y}{N_{\sigma-}} \ ,
\end{eqnarray*}
 $k_b=\sqrt{2m\Delta_b}$, and $\int^{\prime}$ means the integration over states about the Fermi energy, 
i.e., $|\bm{k}|>k_F$. Using the Jacobi-Anger expansion
\begin{eqnarray*}
 e^{i\bm{k}\cdot\bm{r}}=\sum_{m=-\infty}^{+\infty}i^me^{im\theta}J_m(kr) \ ,
\end{eqnarray*}
where $\theta$ is the angle between the two vectors $\bm{k}$ and $\bm{r}$, we find
\begin{equation}
 I_{\sigma}(\bm{r})=\frac{N(0)}{2\pi}\sum_{s=\pm 1}F_{\sigma} \! \left(\bm{r};s\sqrt{k_F^2-k_b^2}\right) ,
 \label{amkondo23}
\end{equation}
where $k_D=\sqrt{2mD}$ and
\begin{eqnarray*}
 F_{\sigma}(\bm{r};b)=\! \int^{\sqrt{k_F^2+k_D^2}}_{k_F} \! dk\frac{J_0(kL_{\sigma}r)}{k-b} \ ,
\end{eqnarray*}
with $|b|<k_F$. An exact evaluation of $F_{\sigma}(\bm{r};b)$ cannot be obtained, we shall determine its 
behaviors in the two regimes: $k_Fr\ll 1$ and $k_Fr\gg 1$.

For $k_Fr\ll 1$, we use the series expansion of $J_0(z)$:
\begin{eqnarray*}
 J_0(z)=1-\frac{1}{4}z^2+O(z^4) \ , 
\end{eqnarray*}
and $F_{\sigma}(\bm{r};b)$ can be written as
\begin{eqnarray*}
 F_{\sigma}(\bm{r};b) \! \! &=& \! \! \! \int^{\sqrt{k_F^2+k_D^2}}_{k_F} \! \frac{dk}{k-b}
 -\frac{(L_{\sigma}r)^2}{4} \! \int^{\sqrt{k_F^2+k_D^2}}_{k_F} \! dk\frac{k^2}{k-b} \\
 \! \! & & \! \! +\cdots \\
 \! \! &=& \! \! \! \left[1-\frac{(bL_{\sigma}r)^2}{4}\right] \! 
 \ln{\! \left(\frac{\sqrt{k_F^2+k_D^2}-b}{k_F-b}\right)} \\
 \! \! & & \! \! -\frac{(k_DL_{\sigma}r)^2}{8}-\frac{b(L_{\sigma}r)^2}{4} \! \left(\sqrt{k_F^2+k_D^2}-k_F
 \right) \\
 \! \! & & \! \! +\cdots \ ,
\end{eqnarray*}
where $\cdots$ denotes the higher order terms. Inserting this form of $F_{\sigma}(\bm{r};b)$ into Eq. 
(\ref{amkondo23}) and noting that $\Delta_b/D,\Delta_b/\epsilon_F\ll 1$, we obtain Eq. (\ref{amkondo15}).

For $k_Fr\gg 1$, we employ the asymptotic expansion of $J_0(z)$:
\begin{eqnarray*}
 J_0(z) &=& \sqrt{\frac{2}{\pi z}}\cos{\! \left(z-\frac{\pi}{4}\right)} \! -\frac{\sqrt{2/\pi}}{8z^{3/2}}
 \sin{\! \left(z-\frac{\pi}{4}\right)} \\
 & & +O(z^{-5/2}) \ ,
\end{eqnarray*}
and $F_{\sigma}(\bm{r};b)$ becomes
\begin{eqnarray*}
 F_{\sigma}(\bm{r};b) \! \! &=& \! \! \sqrt{\frac{2}{\pi L_{\sigma}r}}\! \int^{+\infty}_{k_F} \! dk
 \frac{\cos{(kL_{\sigma}r-\pi/4)}}{\sqrt{k}(k-b)} \\
 \! \! & & \! \! -\frac{\sqrt{2/\pi}}{8(L_{\sigma}r)^{3/2}} \! 
 \int^{+\infty}_{k_F} \! dk\frac{\sin{(kL_{\sigma}r-\pi/4)}}{k^{3/2}(k-b)}+\cdots \\
 \! \! &=& \! \! \frac{2\sqrt{2/\pi}}{\sqrt{L_{\sigma}k_Fr}}\mbox{Re} \! \left[e^{-i\pi/4}G(L_{\sigma}k_Fr)
 \right] \\
 \! \! & & \! \! -\frac{\sqrt{2/\pi}(k_F/b)}{4(L_{\sigma}k_Fr)^{3/2}}\mbox{Im} \! \left[e^{-i\pi/4}G
 (L_{\sigma}k_Fr)\right] \\\
 \! \! & & \! \! +\frac{\sqrt{2/\pi}(k_F/b)}{4(L_{\sigma}k_Fr)^{3/2}}\mbox{Im} \! \left[
 e^{-i\pi/4}\tilde{G}(L_{\sigma}k_Fr)\right] \! +\cdots \ ,
\end{eqnarray*}
where $\cdots$ denotes the higher order terms and
\begin{eqnarray*}
 G(w)=\! \int^{+\infty}_1 \! dt\frac{e^{izt^2}}{t^2-b/k_F} \ , ~~
 \tilde{G}(w)=\! \int^{+\infty}_1 \! dt\frac{e^{izt^2}}{t^2} \ ,
\end{eqnarray*}
with $z=w+i0^+$. In the above, we have set $D\rightarrow+\infty$. The third term is odd in $b$, and thus 
it will not contribute to $I_{\sigma}(\bm{r})$. 

What we are interested in is the behavior of $G(w)$ at $w\gg 1$. In this limit, the integrand oscillates 
quickly such that the contributions cancel, and the integral is supposed to be dominant by the region 
around $t=1$. By making the change of variables $t=1+x$ where $0\leq x\ll 1$, $G(w)$ can be approximated 
as
\begin{eqnarray*}
 G(w)\approx\frac{e^{iw}}{2} \! \int^{+\infty}_0 \! dx\frac{e^{2izx}}{x+(1-b/k_F)/2} \ .
\end{eqnarray*}
Using the identity\cite{Grad}
\begin{eqnarray*}
 \int^{+\infty}_0 \! dx\frac{e^{-\mu x}}{x+\beta}=-e^{\beta\mu}\mbox{Ei}(-\beta\mu) \ ,
\end{eqnarray*}
for $\mbox{Re}\mu>0$, we find
\begin{eqnarray*}
 G(w)=-\frac{e^{iw}}{2}e^{-i(1-b/k_F)z}\mbox{Ei}[i(1-b/k_F)z] \ ,
\end{eqnarray*}
where $\mbox{Ei}(z)$ is the exponential integral function. In terms of the asymptotic expansion\cite{Grad}
\begin{eqnarray*}
 \mbox{Ei}(z)=\frac{e^z}{z} \! \left[1+\frac{1}{z}+O(z^{-2})\right] ,
\end{eqnarray*}
for $|z|\gg 1$, we obtain
\begin{eqnarray*}
 G(w)=\frac{e^{iw}}{2} \! \left[\frac{ik_F}{(k_F-b)w}+\frac{k_F^2}{(k_F-b)^2w^2}+O(w^{-3})\right] .
\end{eqnarray*}
Substituting this result into the expression for $F_{\sigma}(\bm{r};b)$ and summing over $\pm b$, we 
obtain Eq. (\ref{amkondo16}).

%%%%%%%%%%%%%%%%%%%%%%%%%%%%%%%%%%%%%%%%%%%%%%%%%%%%%%%%%%%%%%%%%%%%%%%%%%%%%%%%%%%%%%%%%%%%%%%%%%%%%%%%%
\section{Derivation of Eqs. (\ref{amrkky13}) and (\ref{amrkky14})}\label{a2}

For the AM metal described by $H_0$, the spectral density is written as
\begin{eqnarray*}
 \rho_{\sigma}(\omega,\bm{r})=2\pi \! \int \! \! \frac{d^2k}{(2\pi)^2}e^{i\bm{k}\cdot\bm{r}}\delta
 [\omega-\epsilon_{\sigma}(\bm{k})+\epsilon_F] \ .
\end{eqnarray*}
With the help of the same transformation we used in App. \ref{a1} to treat $\epsilon_{\sigma}(\bm{k})$ 
and the Jacobi-Anger identity to expand the factor $e^{i\bm{k}\cdot\bm{r}}$, we find
\begin{eqnarray*}
 \rho_{\sigma}(\omega,\bm{r})=\frac{m}{\sqrt{1-\nu^2}}\Theta(\omega+\epsilon_F)J_0 \! \left[
 \sqrt{2m(\omega+\epsilon_F)}L_{\sigma}r\right] ,
\end{eqnarray*}
where $\Theta(x)=1,0$ for $x>0$ and $x<0$, respectively. In terms of this result, $W_{ab}(\bm{r})$ can be 
written as
\begin{widetext}
\begin{eqnarray*}
 W_{ab}(\bm{r})=-\frac{m^2}{\lambda_+\lambda_-}\Lambda^{ab}_{\alpha\beta\beta\alpha} \! \int^0_{-\epsilon_F} 
 \! \frac{d\omega_1}{(2\pi)^2}J_0 \! \left[\sqrt{2m(\omega_1+\epsilon_F)}L_{\beta}r\right] 
 \! \int^{+\infty}_0 \! dx \! \left[\frac{J_0(L_{\alpha}rx)}
 {x-\sqrt{2m(\omega_1+\epsilon_F)}}+\frac{J_0(L_{\alpha}rx)}{x+\sqrt{2m(\omega_1+\epsilon_F)}}\right] .
\end{eqnarray*}
\end{widetext}
Using the identity\cite{Grad}
\begin{eqnarray*}
 \int^{+\infty}_0 \! dx\frac{J_0(ax)}{x+k}=\frac{\pi}{2}[\bm{H}_0(ak)-Y_0(ak)] \ ,
\end{eqnarray*}
for $a>0$ and $|\mbox{arg}k|<\pi$ and
\begin{eqnarray*}
 H_0(-z)=-H_0(z) \ , ~~Y_0(-z)=Y_0(z) \ ,
\end{eqnarray*}
where $\bm{H}_{\nu}(z)$ is the Struve function of order $\nu$, we find that
\begin{eqnarray*}
 W_{ab}(\bm{r}) \! \! &=& \! \! \frac{m^2}{\lambda_+\lambda_-}\Lambda^{ab}_{\alpha\beta\beta\alpha} \! 
 \int^0_{-\epsilon_F} \! \frac{d\omega_1}{4\pi}J_0 \! \left[\sqrt{2m(\omega_1+\epsilon_F)}L_{\beta}r
 \right] \\
 \! \! & & \! \! \times Y_0 \! \left[\sqrt{2m(\omega_1+\epsilon_F)}L_{\alpha}r\right] \\
 \! \! &=& \! \! \! \! \frac{mk_F^2\Lambda^{ab}_{\alpha\beta\beta\alpha}}{4\pi\lambda_+\lambda_-} \! \int_0^1 
 \! dxxJ_0(k_FrL_{\beta}x)Y_0(k_FrL_{\alpha}x) \ .
\end{eqnarray*}
Therefore, $\Pi_{ab}(\bm{r})$ is given by
\begin{eqnarray*}
 \Pi_{ab}(\bm{r})=\frac{mk_F^2\Lambda^{ab}_{\alpha\beta\beta\alpha}}{4\pi\lambda_+\lambda_-}
 [I_{\beta\alpha}(\bm{r})+(\alpha\leftrightarrow\beta)] \ .
\end{eqnarray*}
where
\begin{eqnarray*}
 I_{\beta\alpha}(\bm{r})=\! \int_0^1 \! dxxJ_0(k_FrL_{\beta}x)Y_0(k_FrL_{\alpha}x) \ .
\end{eqnarray*}

With the help of the identities
\begin{eqnarray*}
 \int^1_0 \! dxxJ_0(\alpha x)Y_0(\beta x)=\frac{\alpha J_1(\alpha)Y_0(\beta)-\beta J_1(\beta)Y_0(\alpha)}
 {\alpha^2-\beta^2} , 
\end{eqnarray*}
for $\alpha\neq\beta$ and
\begin{eqnarray*}
 \int^1_0 \! dxxJ_0(ax)Y_0(ax)=\frac{1}{2}[J_0(a)Y_0(a)+J_1(a)Y_1(a)] \ , 
\end{eqnarray*}
we find that
\begin{eqnarray*}
 & & I_{\beta\alpha}(\bm{r})+(\alpha\leftrightarrow\beta) \\
 & & =\frac{2[L_{\alpha}J_1(k_FrL_{\alpha})Y_0(k_FrL_{\beta})-(\alpha\leftrightarrow\beta)]}
 {k_Fr(L_{\alpha}^2-L_{\beta}^2)} \ ,
\end{eqnarray*}
for $\alpha\neq\beta$ and 
\begin{eqnarray*}
 & & I_{\beta\alpha}(\bm{r})+(\alpha\leftrightarrow\beta) \\
 & & =J_0(k_FrL_{\alpha})Y_0(k_FrL_{\alpha})+J_1(k_FrL_{\alpha})Y_1(k_FrL_{\alpha}) \ ,
\end{eqnarray*}
for $\alpha=\beta$. On the other hand,
\begin{eqnarray*}
 \Lambda^{ab}_{1111}=\delta_{a3}\delta_{b3}=\Lambda^{ab}_{2222} \ ,
\end{eqnarray*}
and
\begin{eqnarray*}
 \Lambda^{ab}_{1221} &=& \delta_{a1}\delta_{b1}+\delta_{a2}\delta_{b2}+i(\delta_{a1}\delta_{b2}
 -\delta_{a2}\delta_{b1}) \ , \\
 \Lambda^{ab}_{2112} &=& \delta_{a1}\delta_{b1}+\delta_{a2}\delta_{b2}-i(\delta_{a1}\delta_{b2}
 -\delta_{a2}\delta_{b1}) \ .
\end{eqnarray*}
Collecting the above results, we obtain Eqs. (\ref{amrkky13}) and (\ref{amrkky14}).

%%%%%%%%%%%%%%%%%%%%%%%%%%%%%%%%%%%%%%%%%%%%%%%%%%%%%%%%%%%%%%%%%%%%%%%%%%%%%%%%%%%%%%%%%%%%%%%%%%%%%%%%%%%
\section{The slave-boson mean field theory}

By introducing the pseudo-fermion representation of the spin operator $\bm{S}$
\begin{equation}
 S^+=f^{\dagger}_+f_- \ , ~S^-=f^{\dagger}_-f_+ \ , ~
 S_z=\frac{1}{2}\sum_{\sigma=\pm}\sigma f^{\dagger}_{\sigma}f_{\sigma} \ , \label{spin1}
\end{equation}
with the constraint
\begin{equation}
 n_f=\sum_{\sigma=\pm}f^{\dagger}_{\sigma}f_{\sigma}=1 \ . \label{spin11}
\end{equation}
where $f_{\sigma}$ and $f^{\dagger}_{\sigma}$ obey the canonical anticommutation relations, the Kondo 
coupling can be written as
\begin{eqnarray*}
 H_K=J\sum_{\alpha,\beta=\pm}f^{\dagger}_{\alpha}f_{\beta}\psi^{\dagger}_{\beta}(0)\psi_{\alpha}(0)
 +\frac{J}{2}\sum_{\sigma}\psi^{\dagger}_{\sigma}(0)\psi_{\sigma}(0) \ .
\end{eqnarray*}
The second term in $H_K$ describes a local potential scattering, which can be neglected when studying the Kondo
problem. In terms of the pseudo-fermion representation of the spin operator, the partition function is of the 
form
\begin{equation}
 Z=\! \int \! \! D[f^{\dagger}]D[f]D[\psi^{\dagger}]D[\psi]D[\lambda]e^{- \! \int^{\beta}_0 \! d\tau L_K} \ , 
 \label{amkondo4}
\end{equation} 
where
\begin{eqnarray}
 L_K &=& \sum_{\bm{k}\sigma}c^{\dagger}_{\sigma}[\partial_{\tau}+\epsilon_{\sigma}(\bm{k})-\mu]c_{\bm{k}\sigma}
 +\sum_{\sigma}f^{\dagger}_{\sigma}(\partial_{\tau}+i\lambda)f_{\sigma} \nonumber \\
 & & -J\sum_{\alpha,\beta=\pm}f^{\dagger}_{\alpha}\psi_{\alpha}(0)\psi^{\dagger}_{\beta}(0)f_{\beta}-i\lambda 
 \ , \label{amkondo41}
\end{eqnarray}
and $\lambda$ is the Lagrange multiplier to impose the constraint (\ref{spin11}).

To proceed, we perform a Hubbard-Stratonovich transformation on the Kondo coupling term, so that the 
partition function becomes
\begin{equation}
 Z=\! \int \! \! D\mu\exp{\! \left[- \! \int^{\beta}_0 \! d\tau L\right]} \ , \label{amkondo42}
\end{equation}
where $D\mu=D[f^{\dagger}]D[f]D[\psi^{\dagger}]D[\psi]D[\lambda]D[V^{\dagger}]D[V]$ and 
\begin{eqnarray}
 L \! \! &=& \! \! \sum_{\bm{k}\sigma}c^{\dagger}_{\sigma}[\partial_{\tau}+\epsilon_{\sigma}(\bm{k})-\mu]c_{\bm{k}\sigma}
 +\sum_{\sigma}f^{\dagger}_{\sigma}(\partial_{\tau}+i\lambda)f_{\sigma} \nonumber \\
 \! \! & & \! \! +\sum_{\sigma}[V^{\dagger}\psi^{\dagger}_{\sigma}(0)f_{\sigma}+f^{\dagger}_{\sigma}\psi_{\sigma}(0)V]
 +\frac{V^{\dagger}V}{J}-i\lambda \ . ~~ \label{amkondo43}
\end{eqnarray}

A mean-field theory is obtained by neglecting the fluctuations of $V$, $V^{\dagger}$, and $\lambda$. Thus, the
mean-field Hamiltonian is of the form
\begin{eqnarray}
 H_{mf} &=& \sum_{\bm{k}\sigma}c^{\dagger}_{\sigma}[\epsilon_{\sigma}(\bm{k})-\mu]c_{\bm{k}\sigma}
 +\lambda_0(n_f-1)+\frac{|V_0|^2}{J} \nonumber \\
 & & +\frac{1}{\sqrt{\Omega}}\sum_{\bm{k},\sigma}[V_0^*c^{\dagger}_{\bm{k}\sigma}f_{\sigma}+\mathrm{H.c.}] 
 \ , \label{amkondo24}
\end{eqnarray}
where $\lambda_0=i\langle\lambda\rangle$ is a real constant and $V_0=\langle V\rangle$. $H_{mf}$ is quadratic 
in the fermion operators, and thus this problem can be solved exactly.

Instead of diagonalize $H_{mf}$, we would like to determine the Green's functions of $f$-fermions, which is 
defined as
\begin{equation}
 G_{\alpha\beta}(\tau)=-\langle\mathcal{T}_{\tau}\{f_{\alpha}(\tau)f^{\dagger}_{\beta}(0)\}\rangle \ ,
 \label{amkondo25}
\end{equation}
in the imaginary-time formulation. The equation of motion (EOM) for $G_{\alpha\beta}(\tau)$ is
\begin{equation}
 (-\partial_{\tau}-\lambda_0)G_{\alpha\beta}(\tau)=\delta(\tau)\delta_{\alpha\beta}+\frac{V_0}{\sqrt{\Omega}}
 \sum_{\bm{k}}\Gamma_{\alpha\beta}(\tau,\bm{k}) \ , \label{ameom1}
\end{equation}
where
\begin{equation}
 \Gamma_{\alpha\beta}(\tau,\bm{k})=-\langle\mathcal{T}_{\tau}\{c_{\bm{k}\alpha}(\tau)f^{\dagger}_{\beta}(0)\}
 \rangle \ . \label{amkondo26}
\end{equation}
To solve Eq. (\ref{ameom1}), we need the EOM for the function $\Gamma_{\alpha\beta}(\tau,\bm{k})$, which is 
given by
\begin{equation}
 [-\partial_{\tau}-\epsilon_{\alpha}(\bm{k})+\mu]\Gamma_{\alpha\beta}(\tau,\bm{k})=\frac{V^*_0}{\sqrt{\Omega}}
 G_{\alpha\beta}(\tau) \ . \label{ameom11}
\end{equation}

Now Eqs. (\ref{ameom1}) and (\ref{ameom11}) are a closed set of equations. To solve them, we take the Fourier 
transform, yielding 
\begin{eqnarray}
 \tilde{\Gamma}_{\sigma}(i\omega_n,\bm{k}) &=& \frac{V_0^*/\sqrt{\Omega}}{i\omega_n-\epsilon_{\sigma}(\bm{k})+\mu}
 \tilde{G}_{\sigma}(i\omega_n) \ , \nonumber \\
 \tilde{G}_{\sigma}(i\omega_n) &=& \frac{1}{i\omega_n-\lambda_0-\Sigma_{\sigma}(i\omega_n)} \ , 
 \label{amkondo27}
\end{eqnarray}
where $\tilde{G}_{\alpha\beta}=\delta_{\alpha\beta}\tilde{G}_{\alpha}$, 
$\tilde{\Gamma}_{\alpha\beta}=\delta_{\alpha\beta}\tilde{\Gamma}_{\alpha}$, and
\begin{eqnarray*}
 \Sigma_{\sigma}(z) &=& \frac{|V_0|^2}{\Omega}\sum_{\bm{k}}\frac{1}{z-\epsilon_{\sigma}(\bm{k})+\mu} \\
 &=& -|V_0|^2N(0)\ln{\! \left(\frac{D-z}{-D-z}\right)} ,
\end{eqnarray*}
where $D$ is the bandwidth. Taking the analytic continuation $z\rightarrow\omega+i0^+$, we find 
$\Sigma_{\sigma}(\omega)=-i\pi |V_0|^2N(0)$, and thus the Fourier transform of the retarded Green function, 
$\tilde{G}_{R\sigma}(\omega)$, takes the form
\begin{equation}
 \tilde{G}_{R\sigma}(\omega)=\frac{1}{\omega-\lambda_0+i\Delta} \ , \label{amkondo28}
\end{equation}
where $\Delta=\pi |V_0|^2N(0)$. Consequently, the spectral function of the two-point Green function of 
$f$-fermions is
\begin{equation}
 \rho_{\sigma}(\omega)=-2\mbox{Im}\tilde{G}_{R\sigma}(\omega)=\frac{2\Delta}{(\omega-\lambda_0)^2+\Delta^2}
 \ . \label{amkondo3}
\end{equation}

We still have to determine the mean-field values of $|V_0|$ and $\lambda_0$. They are the solutions of the 
mean-field equations:
\begin{equation}
 \frac{V_0^*}{J}=\frac{1}{\sqrt{\Omega}}\sum_{\bm{k},\sigma}\langle c_{\bm{k}\sigma}f^{\dagger}_{\sigma}\rangle
 \ , ~~\langle n_f\rangle=1 \ . \label{ammfe1}
\end{equation}
The first relation in Eq. (\ref{ammfe1}) can be written as
\begin{eqnarray*}
 \frac{V_0^*}{J}=V_0^* \! \int^D_{-D} \! \frac{d\omega}{2\pi}\frac{\tilde{\rho}(\omega)}{e^{-\beta\omega}+1} 
 \ ,
\end{eqnarray*}
where
\begin{eqnarray*}
 \tilde{\rho}(\omega) &\equiv& -2\mbox{Im}\frac{1}{\sqrt{\Omega}V_0^*}\sum_{\bm{k},\sigma}
 \tilde{\Gamma}_{R\sigma}(\omega,\bm{k}) \\
 &=& \frac{4\pi N(0)(\omega-\lambda_0)}{(\omega-\lambda_0)^2+\Delta^2} \ .
\end{eqnarray*}
At $T=0$, this equation reduces to
\begin{eqnarray*}
 \frac{1}{J}=\! \int_0^D \! \frac{d\omega}{2\pi}\tilde{\rho}(\omega)=2N(0) \! \int_0^D \! d\omega
 \frac{\omega-\lambda_0}{(\omega-\lambda_0)^2+\Delta^2} \ ,
\end{eqnarray*}
for $|V_0|\neq 0$. Assuming that $D\gg|\lambda_0|,\Delta$ and performing the frequency integral, the above 
equation becomes
\begin{equation}
 \frac{1}{J}=N(0)\ln{\! \left(\frac{D^2}{\lambda_0^2+\Delta^2}\right)} . \label{ammfe11}
\end{equation}
It is clear that Eq. (\ref{ammfe11}) has solutions only when $J>0$.

The second relation in Eq. (\ref{ammfe1}) can be written as
\begin{eqnarray*}
 1=\sum_{\sigma} \! \! \int^D_{-D} \! \frac{d\omega}{2\pi}\frac{\rho_{\sigma}(\omega)}{e^{\beta\omega}+1} 
 \ .
\end{eqnarray*}
At $T=0$, this equation reduces to
\begin{eqnarray*}
 1=\sum_{\sigma} \! \int^0_{-D} \! \frac{d\omega}{2\pi}\rho_{\sigma}(\omega)=\frac{2\Delta}{\pi} \! 
 \int^0_{-D} \! \frac{d\omega}{(\omega-\lambda_0)^2+\Delta^2}
\end{eqnarray*}
Performing the frequency integral and taking $D\rightarrow+\infty$, the above equation becomes
\begin{equation}
 \tan^{-1}{\! \left(\frac{\lambda_0}{\Delta}\right)} \! =0  \ . \label{ammfe12}
\end{equation}
The solution of Eq. (\ref{ammfe12}) is $\lambda_0=0$. Inserting this result into Eq. (\ref{ammfe11}), we get
\begin{equation}
 \Delta=De^{-1/[2N(0)J]} \ . \label{amkondo33}
\end{equation}
If we identify $T_K=\Delta$, then Eq. (\ref{amkondo33}) is identical to the one obtained from the one-loop RG
equation.

%\end{appendices}
%%%%%%%%%%%%%%%%%%%%%%%%%%%%%%%%%%%%%%%%%%%%%%%%%%%%%%%%%%%%%%%%%%%%%%%%%%%%%%%%%%%%%%%%%%%%%%%%%%%%%%%%%%%

\end{document}